\documentclass[aps,pb,amsmath,amssymb,twocolumn,twoside,showpacs,superscriptaddress,nofootinbib,longbibliography]{revtex4-2}
\usepackage[pdftex]{graphicx}
\usepackage[colorlinks=true,citecolor=blue]{hyperref}
\usepackage{graphicx,textcomp,dcolumn,hyperref,upgreek}

\usepackage{amssymb}
\usepackage{amsmath}
\usepackage{xcolor}

\usepackage{float}

\usepackage{dsfont}

\usepackage{physics}

\usepackage{xspace}

\usepackage[english]{babel}
\usepackage{lmodern}

\usepackage{blindtext}

\newcommand{\z}{_{\noindent z}}
\newcommand{\TX}{{\cal T}}

\begin{document}

\title{
Unraveling  spin dynamics from charge fluctuations
}
\author{Eric~Kleinherbers}
\email{eric.kleinherbers@uni-due.de}
\affiliation{Faculty of Physics and CENIDE, University of Duisburg-Essen, 47057 Duisburg, Germany}
\affiliation{Department of Physics and Astronomy, University of California, Los Angeles, California 90095, USA}

\author{Hendrik~Mannel}
\affiliation{Faculty of Physics and CENIDE, University of Duisburg-Essen, 47057 Duisburg, Germany}

\author{Jens~Kerski}
\affiliation{Faculty of Physics and CENIDE, University of Duisburg-Essen, 47057 Duisburg, Germany}

\author{Martin~Geller}
\affiliation{Faculty of Physics and CENIDE, University of Duisburg-Essen, 47057 Duisburg, Germany}

\author{Axel~Lorke}
\affiliation{Faculty of Physics and CENIDE, University of Duisburg-Essen, 47057 Duisburg, Germany}

\author{J{\"u}rgen~K{\"o}nig}
\affiliation{Faculty of Physics and CENIDE, University of Duisburg-Essen, 47057 Duisburg, Germany}  
                              
\date{\today}

\begin{abstract}
The use of single electron spins in quantum dots as qubits requires detailed knowledge about the processes involved in their initialization and operation as well as their relaxation and decoherence.
In optical schemes for such spin qubits, spin-flip Raman as well as Auger processes play an important role, in addition to environment-induced spin relaxation.
In this paper, we demonstrate how to quantitatively access all the \textit{spin}-related processes in one go by monitoring the \textit{charge} fluctuations of the quantum dot.
For this, we employ resonance fluorescence and analyze the charge fluctuations in terms of waiting-time distributions and full counting statistics characterized by factorial cumulants.
\end{abstract}

\maketitle
\section{Introduction}

The spin of a single electron in a quantum dot is a potential candidate for a quantum bit (qubit) in quantum-computation schemes~\cite{loss_1998,elzerman_2004,amasha_2008,ladd_2010}.
Elementary operations on the spin state can be realized electrically in lithography-defined quantum dots by fast control of the exchange interaction~\cite{petta_2005}.
In self-assembled quantum dots, the optical transitions can be used to initialize and control the spin by an induced laser field~\cite{imamoglu_1999,xu_2007,press_2008,muller_2013,gao_2015} and connect the stationary spin qubit with the flying photon qubit by a spin-photon interface~\cite{yilmaz_2010, degreve_2012}. 
These optical transitions include spin-flip Raman scattering, a process in which the quantum-dot spin is reversed by optically exciting a trion, a many-body state consisting of two electrons and a hole, as an intermediate state.
Spin-flip Raman scattering~\cite{debus_2014_2}, thus, provides a possibility of optical spin pumping from the ground to the excited spin state~\cite{dreiser_2008,atatuere_2006}.

\begin{figure}[ht]
  \begin{center}
   \includegraphics[width=.45\textwidth]{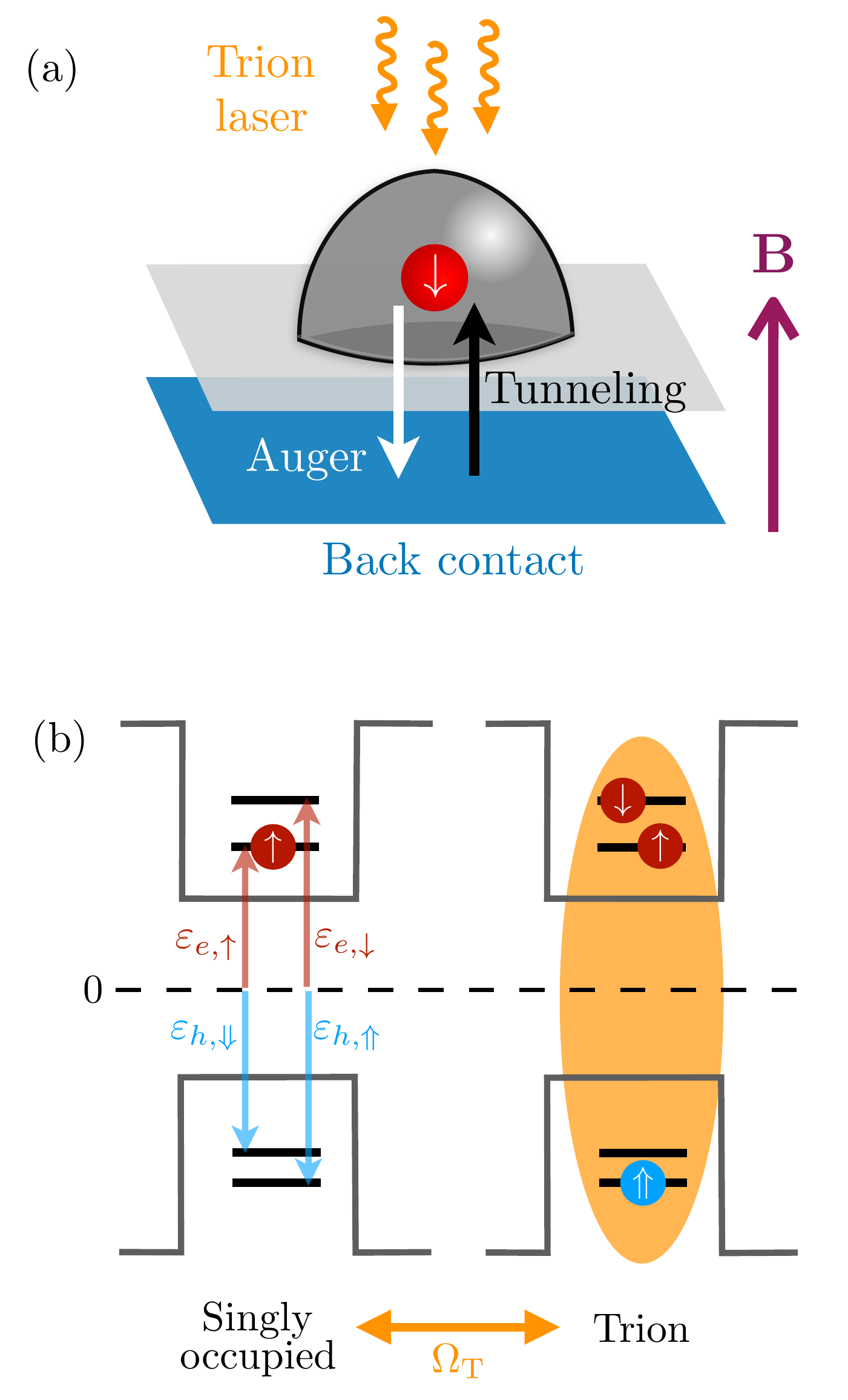}
  \end{center}
  \caption{(a) Self-asssembled  InAs quantum dot in an applied magnetic field (purple) and subject to a laser field (orange) that resonantly drives a trion. Charge fluctuations into the back contact (blue) are possible via tunneling (black) and the Auger process (white).
  (b) A laser field drives a trion transition in the quantum dot.}
  \label{fig:auger_exp_setup}
\end{figure}

The design of protocols for spin-qubit operations requires an accurate knowledge of all the rates of the processes that affect the quantum-dot spin, as well as their dependence on control parameters such as the strength of a static magnetic field or the intensity of an applied laser field.
In this paper, we study a single electron in a self-assembled InAs quantum dot in an applied magnetic field and subject to a laser field, see Fig.~\ref{fig:auger_exp_setup}(a) ~\cite{mannel_2021,kurzmann2019optical}.
The quantum-dot charge state is tuned by a gate voltage in such a way that the quantum dot is predominantly singly occupied to define a spin qubit~\cite{lochner_2020}. 

There are four relevant processes that affect the spin dynamics in our system.
The first one is \textit{spin relaxation}~\cite{gillard_2021}.
It is the consequence of the coupling of the electron spin to mobile electrons in the back contact (blue in Fig.~\ref{fig:auger_exp_setup}(a)) via cotunneling~\cite{dreiser_2008}, nuclear spins in the quantum dot by hyperfine interaction~\cite{dahbashi_2014,merkulov_2002,bracker_2005,kuhlmann_2013}, and/or phonons by spin-orbit interaction~\cite{khaetskii_2001}.
Second, a laser properly tuned to a trion resonance, see Fig.~\ref{fig:auger_exp_setup}(b),  allows for optical spin pumping, referred to as \textit{spin-flip Raman} process~\cite{atatuere_2006,dreiser_2008,debus_2014_2,mannel_2021}.
While in this process, the decay of the trion state is accompanied with the emission of a photon, there is another decay channel for the trion, in which the energy gained from the recombination of the electron-hole pair is carried away by knocking the remaining electron out of the quantum dot into states high in the conduction band, from where it can relax down into the  back contact (white arrow). 
This third process is called \textit{Auger recombination} and has only recently been recognized as a relevant transition in a single self-assembled quantum dot~\cite{kurzmann_2016_auger,lobl_2020,lochner_2020,gillard_2021}.
The Auger recombination we consider here is nonradiative (in contrast to Ref.~\cite{lobl_2020}) and leaves the quantum dot in the empty state. 
Thereby, not only the quantum-dot charge but also its spin has gone.
Both the rates for spin-flip Raman and for Auger processes depend on the intensity of the laser driving the trion transition.
Finally, the fourth process is \textit{tunneling} of an electron from the back contact into the quantum dot (black arrow).
This tunneling process also affects the spin dynamics, although only indirectly in combination with the Auger process~\cite{mannel_2021}.

How can one measure these four rates?
A seemingly natural strategy would be to determine these four rates separately in a pump-probe manner.
After preparing the corresponding initial states at a well-defined time, one measures the decay into the corresponding final state as a function of time. 
Such an approach, however, requires the possibility to properly identify and discriminate the different quantum-dot states in a sophisticated optical experiment~\cite{gillard_2021}.
Furthermore, more than one of the four processes are, in general, present at the same time, which makes it difficult to disentangle the individual contributions.

The main purpose of this paper is to suggest an alternative way of determining the four spin-related rates.
It differs from the strategy outlined above in several respects.
First, even though we aim at addressing the \textit{spin} dynamics we only detect the \textit{charge} dynamics of the quantum dot.
This makes the requirement to directly discriminate between different spin states dispensable.
Second, although we want to address dynamics, we can restrict ourselves to perform a steady-state measurement but keep the time information by recording the full time trace of individual quantum jumps (tunneling-in and Auger events).
Since tunneling is a stochastic phenomenon, the charge state strongly fluctuates in time. 
The individual quantum jumps are, however, not independent of each other, which opens the possibility to determine the desired quantities from the analysis of the charge fluctuations.
Third, by analyzing the charge fluctuations we can, as we will show below, address all four rates in one go.
To achieve this, the analysis of waiting times~\cite{brandes_2008,albert_2012,rudge_2018,kleinherbers_2021_synchro} as well as the full counting statistics~\cite{levitov_1993,blanter_2000,marcos_2010,rudge_2019} in terms of so-called factorial cumulants~\cite{beenakker_counting_2001,flindt2009universal, kambly_2011,stegmann2015detection,stegmann2016short,kleinherbers2018revealing,koenig_2021,kleinherbers_2021_pushing} will be used.

This paper is organized as follows.
In Sec.~\ref{sec:auger_system}, we describe the quantum-dot states and the transitions between them.
Then, in Sec.~\ref{sec:auger_fcs}, we use waiting-time distributions and full counting statistics to determine all the transition rates that fully describe the dynamics of the system.
All this can be done by statistically analyzing a single telegraph signal stream. 
Besides the charge fluctuations due to the Auger process and electron tunneling, we can infer the internal spin dynamics of the system due to spin relaxation and optical spin pumping. 
Finally, in Sec.~\ref{sec:auger_conclusion}, we conclude our findings.

\section{Quantum-dot states and transitions} \label{sec:auger_system}

The self-assembled InAs quantum dot studied here is the same as in Ref.~\cite{mannel_2021}.
We use time-resolved resonance fluorescence on a single quantum dot~\cite{vamivakas_2010,matthiesen_2013, kurzmann_2016,kurzmann2019optical}, where the quantum dot layer is embedded in a
p-i-n diode structure with a highly n-doped layer as the charge reservoir and a highly p-doped layer as the epitaxial top gate~\cite{lochner_2019}. 
The gate voltage is tuned such that only two charge states of the quantum dot play a role.
The quantum dot can either be charge neutral or charged with one extra electron of spin up or down, residing in a conduction-band state. 
An applied laser field drives the quantum dot into a trion state, but it also gives rise to both spin-flip Raman scattering and Auger recombination.
In addition to this \textit{trion laser}, we apply another one with a frequency adjusted to the exciton resonance, which we refer to as the \textit{exciton laser}.
While the trion laser generates the spin dynamics we want to study, the exciton laser (which was not included in Ref.~\cite{mannel_2021}) is used to measure the charge dynamics.

\subsection{Hamiltonian}

To include all relevant states of the quantum dot, we use the model Hamiltonian 
\begin{align}
H_\text{S}=&\sum_{\sigma} \varepsilon_{e,\sigma} e_\sigma^\dagger e_\sigma^{\phantom{\dagger}} +\sum_{\tau} \varepsilon_{h,\tau} h_\tau^\dagger h_\tau^{\phantom{\dagger}}+ U_{ee} n_{e,\uparrow} n_{e,\downarrow} \\ &+U_{hh} n_{h,\Uparrow} n_{h,\Downarrow} \nonumber - U_{eh} \left(n_{e,\uparrow}+n_{e,\downarrow}\right)\left(n_{h,\Uparrow}+n_{h,\Downarrow} \right),
\end{align}
where $e_\sigma^\dagger$ and $e_\sigma^{\phantom{\dagger}}$ describe the creation and annihilation operators of spin-$1/2$ electrons with $\sigma\in\{\uparrow, \downarrow\}$. In addition,  $h_\tau^\dagger$ and  $h_\tau^{\phantom{\dagger}}$ create and annihilate heavy holes with angular momentum $3/2$ and $z$-component $\pm 3/2$ which we write (following Ref.~\cite{dreiser_2008}) as $\tau \in \{ \Uparrow, \Downarrow\}$.
The associated light holes with $z$-component $\pm 1/2$ are typically several tens of $\text{meV}$ higher in energy and can be neglected~\cite{bayer_2002}. Thus, the eigenstates of the quantum system can be labelled by $\ket{\chi_e,\chi_h}$ with $\chi_e\in\{0,\uparrow,\downarrow,\uparrow\downarrow\}$ and $\chi_h\in\{0,\Uparrow,\Downarrow,\Uparrow\Downarrow\}$ denoting  zero, single, or double occupation of the quantum dot with electron and holes, respectively. 

\begin{figure}[t]
	\centering
    \includegraphics[width=.45\textwidth]{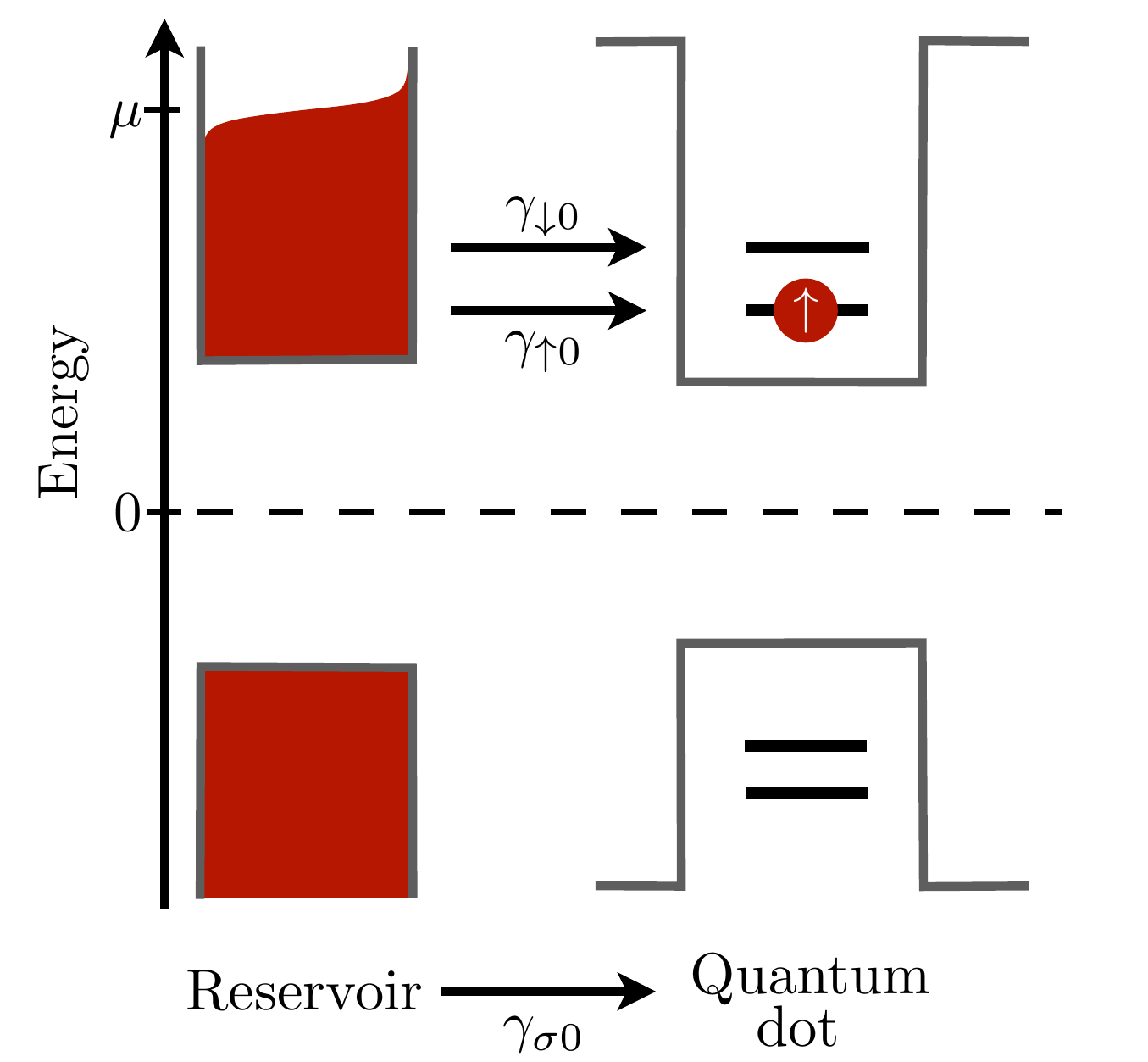}
		\caption{Quantum dot coupled to an electron reservoir which enables tunneling-in events.} 
	\label{fig:auger_coupling}
\end{figure}

The first two terms of the Hamiltonian $H_\text{S}$ describe the single-particle energies $\varepsilon_{e,\downarrow}>\varepsilon_{e,\uparrow}>0$ and $\varepsilon_{h,\Uparrow}>\varepsilon_{h,\Downarrow}>0$ to add a single electron and hole to the system, see Fig.~\ref{fig:auger_exp_setup}(b), where we choose the dashed line as the reference for zero energy.
A static magnetic field of $B=4\,\text{T}$ in the \textit{Faraday configuration} (in growth direction, parallel to the laser beam) 
lifts the degeneracy of the spin states, giving rise to Zeeman splittings $\Delta_e=\varepsilon_{e,\downarrow}{-}\varepsilon_{e,\uparrow}>0$ and $\Delta_h=\varepsilon_{h,\Uparrow}{-}\varepsilon_{h,\Downarrow}>0$.
They are related to the magnetic field $B$ via $\Delta_{e/h}=\vert g_{e/h} \mu_\text{B} B \vert$ with the Bohr magneton $\mu_\text{B}$ and $g$-factors of about $g_e=-0.8$ and $g_h=0.2$~\cite{debus_2014_2,mannel_2021}.
The remaining three terms model many-body interactions. 
When either two electrons or two holes occupy the quantum dot, the repulsive Coulomb interaction with charging energy $U_{ee}>0$ and $U_{hh}>0$ must be paid, respectively. 
In contrast, electrons and holes attract each other, which gives rise to an energy gain $U_{eh}>0$. 

As indicated in Fig.~\ref{fig:auger_coupling}, the quantum dot is tunnel coupled to an electron reservoir (red) which is characterized by the temperature $T$ and the electrochemical potential $\mu$.
The system is tuned such that $\mu{-}\varepsilon_{e,\sigma} \gg k_\text{B} T$ and $\varepsilon_{e,\sigma}{+}U_{ee}{-}\mu \gg k_\text{B} T$, i.e., the quantum dot is almost exclusively occupied by one electron.
A second electron is prohibited by a strong Coulomb repulsion $U_{ee}$ between the electrons. 
Hence, whenever the quantum dot is empty, electrons with spin $\sigma$ can tunnel into the system with rate $\gamma_{\sigma 0}$. 
The reciprocal process, tunneling out of an electron with spin $\sigma$, is exponentially suppressed and, therefore, neglected.

In addition, the trion laser resonantly drives an optical transition whenever the system is in the energetically lower spin-up state $\ket{\uparrow,0}$, see Fig.~\ref{fig:auger_exp_setup}(b).  
The optical excitation lifts an electron across the band gap, creating in total two electrons and a hole in an interacting many-body state called trion (indicated by the shaded orange region). 
For a dipole transition, the selection rules only allow for a change of angular momentum by $\pm 1$.
The trion resonance is, therefore, described by
\begin{align}
\ket{\uparrow,0}=\ket{\uparrow} \leftrightarrow\ket{\text{T}}= \ket{\uparrow\downarrow,\Uparrow}, 
\end{align}
with angular momentum change of $+1$ and excitation frequency $\omega_{\text{T}}=E_{\text{T}}{-}E_{\uparrow}=\varepsilon_{e,\downarrow}+\varepsilon_{h,\Uparrow}-2U_{eh}+U_{ee}$.
We note here that, in principle, there is another trion resonance, described by $\ket{\downarrow,0} \leftrightarrow \ket{\uparrow\downarrow,\Downarrow}$ with angular momentum change $-1$ and excitation frequency $\omega_{\text{T}}+(g_e-g_h)\mu_\text{B}B$.
As a consequence of $g_e-g_h<0$, the excitation frequency for this second transition is smaller than the first one introduced above.
Due to their difference in frequency, we refer to them as red (lower frequency) and blue (higher frequency) trion, respectively~\cite{mannel_2021}.
The small linewidth of the trion laser allows us to tune it such that only the blue trion transition is driven and the red trion does not play any role.

Not relevant for the spin dynamics but important for the charge-detection scheme we use in our experiment is a second (exciton) laser that drives the transition 
\begin{align}
    \ket{0,0}=\ket{0} \leftrightarrow\ket{\text{X}}= \ket{\downarrow,\Uparrow}
\end{align}
with angular momentum change of $+1$ and excitation frequency $\omega_{\text{X}}=\varepsilon_{e,\downarrow}+\varepsilon_{h,\Uparrow}-U_{eh}$.
Similarly to the trions, there are also two excitons. 
In addition to the blue exciton introduced above, there is a red exciton, described by $\ket{0} \leftrightarrow \ket{\uparrow,\Downarrow}$ with angular momentum change $-1$ and excitation frequency $\omega_{\text{X}}+(g_e-g_h)\mu_\text{B}B$.
We tune, however, the exciton laser such that only the blue exciton is driven.
We comment that, in general, there is also a fine structure splitting of the bright excitons even in the absence of a magnetic field, $B=0$, when the quantum dot deviates from a perfectly circular shape~\cite{bayer_2002}. 
For simplicity, we neglect this splitting in the Hamiltonian.

We remark here in passing that for the quantum dot under consideration, the (blue) trion frequency turns out to be smaller than the (blue) exciton frequency $\omega_\text{X}>\omega_\text{T}$.
This statement is equivalent to the fact that the electron-hole attraction is larger than the electron-electron repulsion, $U_{eh}>U_{ee}$.

To summarize, there are five relevant quantum-dot states, as illustrated in  Fig.~\ref{fig:auger_states}(a).
Two of them, $\ket{0}$ and $\ket{\text{X}}$, are charge neutral.
The remaining three, $\ket{\uparrow}$, $\ket{\downarrow}$, and $\ket{\text{T}}$ carry one negative elementary charge.
These are the states being involved in spin-qubit operations.
Since it is difficult to access the spin degree of freedom directly, we use the transition to/from the charge neutral state to monitor charge fluctuations, from which we unravel the spin dynamics sketched in Fig.~\ref{fig:auger_states}(a).
To capture the full dynamics of the system, we need to specify all the transitions between the different quantum-dot states.

\subsection{Laser-independent transitions}
There are two transitions that occur already in the absence of laser fields.

\subsubsection{Tunneling}
First, there is tunneling between quantum dot and back contact. 
For the gate voltage chosen in this experiment, tunneling of an electron with spin $\sigma\in\{\uparrow,\downarrow\}$ from the back contact to the quantum dot occurs with rate $\gamma_{\sigma 0}$, while the opposite process is exponentially suppressed.
As a consequence, in the absence of laser driving, the quantum dot would almost exclusively be occupied with a single electron.

\subsubsection{Spin relaxation}

The spin of the quantum-dot electron can relax with rate $\gamma_{\uparrow\downarrow}$ from the excited state $\ket{\downarrow}$ to the ground state $\ket{\uparrow}$.
The inverse process is also possible but suppressed by the Boltzmann factor, $\gamma_{\downarrow\uparrow}=e^{-\Delta_e/(k_\text{B} T)}\gamma_{\uparrow\downarrow}$, which involves the ratio of the Zeeman energy $\Delta_e$ and temperature.
Our experiment is performed at temperature $T=4.2 \, \text{K}$ in a magnetic field of $B=4 \, \text{T}$, which yields $\gamma_{\downarrow\uparrow} = 0.6\,  \gamma_{\uparrow\downarrow}$. 
Possible mechanisms to change the electron spin are hyperfine interaction with the surrounding nuclear spins, tunnel coupling to the electron reservoir, and coupling to phonons mediated by spin-orbit interaction. 
For high magnetic fields ($>2\,\text{T}$) the latter process is the dominant one~\cite{kroutvar_2004,dreiser_2008}.

\subsection{Trion-laser driven transitions}
We now consider the transitions induced by the trion laser.

\subsubsection{Trion excitation}
The trion laser coherently drives a (blue) trion transition, $\ket{\uparrow}\leftrightarrow\ket{\text{T}}$.
This is indicated with an orange arrow $\Omega_\text{T}$ in Fig.~\ref{fig:auger_states}(a).
The respective coupling strength is given by the Rabi frequency $\Omega_\text{T}$ which can be tuned by the laser \textit{intensity} $I_\text{T}\propto\Omega_\text{T}^2$ (while keeping the laser \textit{frequency} $\omega_\text{T}$ constant).
In addition to coherent driving, which includes both photon absorption and \textit{stimulated} emission, the excited trion state $\ket{\text{T}}$ can decay back to the state $\ket{\uparrow}$ by \textit{spontaneous} emission of a photon at the rate $\gamma_{\uparrow \text{T}}$ (black arrow $\gamma_{\uparrow \text{T}}$ in Fig.~\ref{fig:auger_states}(a)).

\subsubsection{Spin-flip Raman scattering}
The trion can not only relax back to the state $\ket{\uparrow}$ it was excited from.
There is another decay channel, namely towards the state $\ket{\downarrow}$.
This process is referred to as spontaneous spin-flip Raman scattering since it flips the spin of the quantum-dot electron and the frequency $\omega_{\text{T}}-\Delta_e$ of the emitted photon is reduced by the Zeeman energy $\Delta_e$ from the frequency $\omega_{\text{T}}$ of the absorbed photon~\cite{nesbet_1971,debus_2014}.
Spin-flip Raman scattering leads to optical spin pumping~\cite{shabaev_2003}, which can even give rise to a population inversion of the spin states.

\begin{figure*}[t]
  \begin{center}
    \includegraphics[width=.95\textwidth]{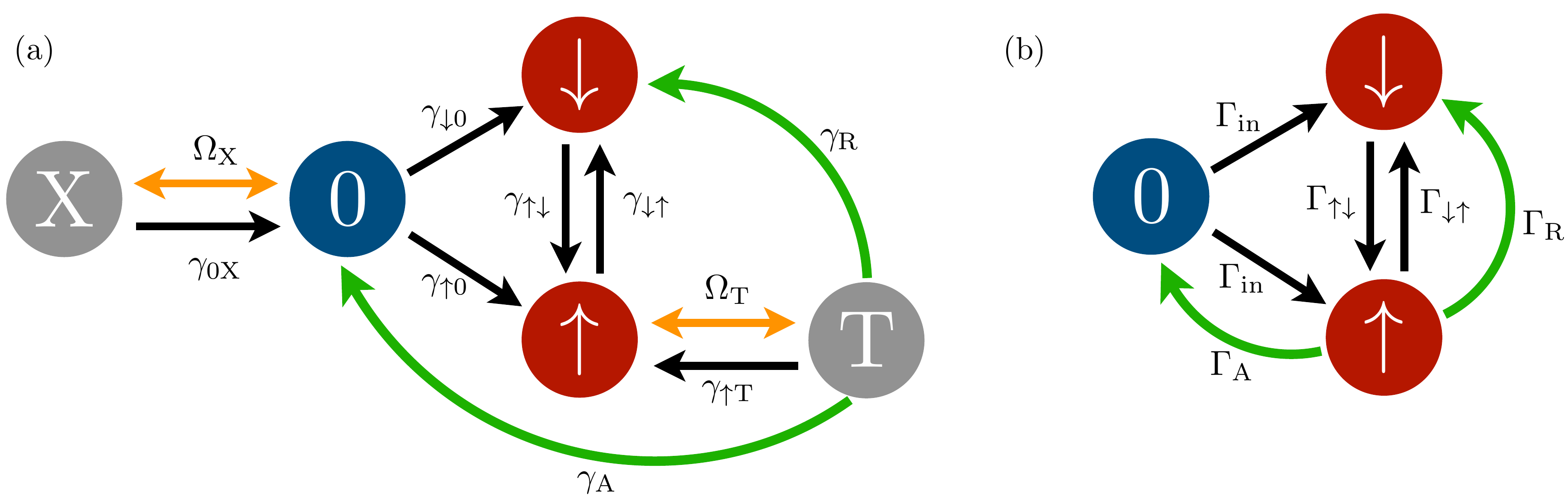}
  \end{center}
  \vspace{-20pt}
  \caption{Relevant states and transitions for the quantum-dot system. (a) Full system and (b) effective three-state system.}\label{fig:auger_states}
\end{figure*}

Phenomenologically, the process is described by the relaxation rate $\gamma_\text{R}$ from the trion to spin-down state, see Fig.~\ref{fig:auger_states}(a), which is energetically higher than the spin-up state.
The rate is much smaller than the spin-conserved spontaneous decay, $\gamma_\text{R}\ll\gamma_{\uparrow\text{T}}$.
This is a consequence of the fact that the spin-flip decay process, $\ket{\uparrow\downarrow,\Uparrow}\rightarrow \ket{\downarrow}$, which changes the $z$-component of the angular momentum by $-2$, is forbidden by the optical selection rules. 
Thus, an additional process is needed to flip the spin~\cite{calarco_2003}. 
 
We review two relevant mechanisms~\cite{dreiser_2008,lu_2010}.
For small magnetic fields, the predominant mechanism is the hyperfine interaction with the nuclear spins which can be described by an effective fluctuating magnetic field called Overhauser field~\cite{hanson_2007}. 
Its transversal component $B_\perp$ (perpendicular to the externally applied magnetic field $B$) mixes the spin states $\ket{\uparrow}$ and $\ket{\downarrow}$ leading to the effective eigenstate $\ket{\downarrow,0}_\text{mix}\sim\ket{\downarrow,0}+\alpha_e \ket{\uparrow,0}$, where $\alpha_e \sim{B_\perp}/{B}$ with $|\alpha_e|\ll 1$ describes the mixing. 
Due to the admixture, the optical selection rules are relaxed and an effective spin-flip process $\ket{\uparrow\downarrow,\Uparrow}\rightarrow\ket{\downarrow,0}_\text{mix}$ is allowed because $\ket{\uparrow\downarrow,\Uparrow} \rightarrow \alpha_e \ket{\uparrow,0}$ changes the $z$-component of the angular momentum by $-1$ only.  
Therefore, the spin-flip Raman scattering process becomes possible  with a rate given by $\gamma_\text{R}\sim \vert\alpha_e\vert^2\gamma_{\uparrow\text{T}}$.

For large magnetic fields, the dominant mechanism is due to mixing between heavy and light holes~\cite{dreiser_2008,calarco_2003}.
In particular, the effective heavy-hole state participating in the optical transition has a small contribution from the spin-up light hole, $\ket{0,\Uparrow}_\text{mix}\sim \ket{0,\Uparrow}+\alpha_h \ket{0,\uparrow}$, where $|\alpha_h| \ll 1$. 
Again, the optical selection rules are relaxed and the transition $\ket{\uparrow\downarrow,\Uparrow}_\text{mix}\rightarrow\ket{\downarrow,0}$ is allowed since $\alpha_h \ket{\uparrow\downarrow,\uparrow}\rightarrow\ket{\downarrow,0}$ changes the $z$-component of the angular momentum by $-1$. 
Hence, the heavy-hole/light-hole mixing enables the spin-flip Raman scattering with the rate $\gamma_\text{R} \sim \vert \alpha_h\vert^2  \gamma_{\uparrow\text{T}}$. 

\subsubsection{Auger recombination}

The trion-laser field does not only influence the quantum-dot spin via spin-flip Raman scattering, it also affects the charge dynamics via Auger recombination.
The Auger effect, known from atomic physics, describes the nonradiative decay of an excited atom in which the excitation energy is used to eject one or more electrons from the atom instead of emitting a photon. 
A similar effect has been found in colloidal quantum dots~\cite{klimov_2000} and, more recently, in self-assembled quantum dots~\cite{kurzmann_2016_auger}. 
In a bulk semiconductor, the Auger effect is suppressed because the particles involved must exactly satisfy the constraints imposed by energy and momentum conservation. However, due to quantum confinement and the resulting momentum uncertainty, these constraints are relaxed in quantum dots and the Auger effect can have a significant impact. 
For example, it can reduce the photon emission~\cite{pietryga_2008}.

Here, the Auger process is realized as an alternative decay channel of the trion state. Instead of spontaneously emitting a photon, the trion can decay nonradiatively by transferring the recombination energy $\omega_{\text{T}}$
of the electron and hole to the second electron which is knocked out of the quantum dot into the back contact. The respective transition rate is denoted as $\gamma_\text{A}$ (see also Fig.~\ref{fig:auger_states}(a)).

With each Auger process, an electron is ejected from the quantum dot, leaving it in the empty state $\ket{0}$. 
Only after a certain time, a new electron of spin $\sigma$ tunnels in at the rate $\gamma_{\sigma0}$ via the back contact, where $\sigma\in\{\uparrow,\downarrow\}$, see Fig.~\ref{fig:auger_states}(a). 
Without Auger recombination, the quantum dot would be, for the chosen gate voltage, always singly occupied and no charge fluctuations would occur.

\subsection{Exciton-laser driven transition}
We monitor as a function of time the quantum-dot charge by using an optical readout scheme based on the excitation of excitons~\cite{kurzmann_2016_auger,kurzmann_2016,lochner_2020}. 
The latter are generated by a second (exciton) laser with fixed laser intensity $I_\text{X}$ (and thus fixed Rabi frequency $\Omega_\text{X}$).
The laser frequency is tuned to the (blue) exciton resonance, $\ket{0}\leftrightarrow \ket{\text{X}}$.
In addition to coherent driving by the laser field, the exciton $\ket{\text{X}}$ can decay back to the state $\ket{0}$ by \textit{spontaneous} emission of a photon at the rate $\gamma_{0 \text{X}}$.
We note that, in principle, the trion transition can also be used for an optical readout. However, for this quantum dot, the photon yield of the trion transition is too low.

\subsection{Effective three-state model}

The optical transitions described by $\Omega_\text{X},\Omega_\text{T},\gamma_{0\text{X}},\gamma_{\uparrow\text{T}}$ (as well as $\gamma_\text{A},\gamma_\text{R}$) 
are much faster than the remaining transitions.
This allows us to effectively eliminate the exciton and trion state from the dynamics, see Fig.~\ref{fig:auger_states}(b), by subsuming state $\ket{\text{X}}$ into $\ket{0}$ and state $\ket{\text{T}}$ into $\ket{\uparrow}$.
To be consistent, we use effective transition rates, which depend, in general, on the laser intensities.
For weak driving of the exciton resonance, $\Omega_\text{X}\ll \gamma_{0\text{X}}$, as is the case in our experiment, the charge-neutral dot is almost always in state $\ket{0}$ and not in state $\ket{\text{X}}$, and the rate for tunneling in needs not be renormalized~\cite{kurzmann_2016}.
The same holds true for the spin-flip relaxation from the ground to the excited spin state, since also the trion transition is only weakly driven.

The situation is different for spin-flip Raman and Auger processes.
They are only possible when the charged quantum dot is in the trion state.
Therefore, the spin-flip Raman and Auger rates need to be multiplied with the degree of saturation $n_{\text{T}}$ of the trion state, i.e., the probability to be in state $\ket{\text{T}}$, normalized by the sum of the probabilities for $\ket{\uparrow}$ and $\ket{\text{T}}$.
For weak driving, this degree of saturation is given by $n_\text{T}= \Omega_\text{T}^2/\gamma_{\uparrow\text{T}}^2\ll 1$.

This leads to the four effective rates
\begin{align}\label{eq:auger_effrates}
\Gamma_\text{in}= \gamma_{\sigma 0},  \quad  \Gamma_{\uparrow\downarrow}= \gamma_{\uparrow\downarrow},   \quad \Gamma_\text{A}=n_\text{T} \gamma_\text{A}, \quad \Gamma_\text{R}=n_\text{T} \gamma_\text{R},
\end{align}
together with the inverse spin relaxation rate  $\Gamma_{\downarrow\uparrow}= e^{-\Delta_e/(k_\text{B} T)}\Gamma_{\uparrow\downarrow}$, see Fig.~\ref{fig:auger_states}(b). Note that we implicitly assumed spin-independent tunneling rates, $\gamma_{\uparrow 0}=\gamma_{\downarrow 0}$, which is justified for $\mu-\varepsilon_e \gg k_\text{B}T$ ensured by the chosen gate voltage, see Fig.~\ref{fig:auger_coupling}.
The relative size of the four rates in Eq.~\eqref{eq:auger_effrates} depends on the different microscopic details of the underlying processes and is therefore, a priori, unknown.
It is the aim of our analysis to determine them.
On the other hand, Eq.~(\ref{eq:auger_effrates}) predicts how the rates depend on the trion laser intensity in the weak-driving regime.
An increase of the trion laser intensity is simply taken into account by multiplying the spin-flip Raman and the Auger rates with the same factor (the degree of saturation $n_\text{T}$ of the trion) while keeping all other rates unchanged. 

\begin{figure*}[t]
  \includegraphics[width=1\textwidth]{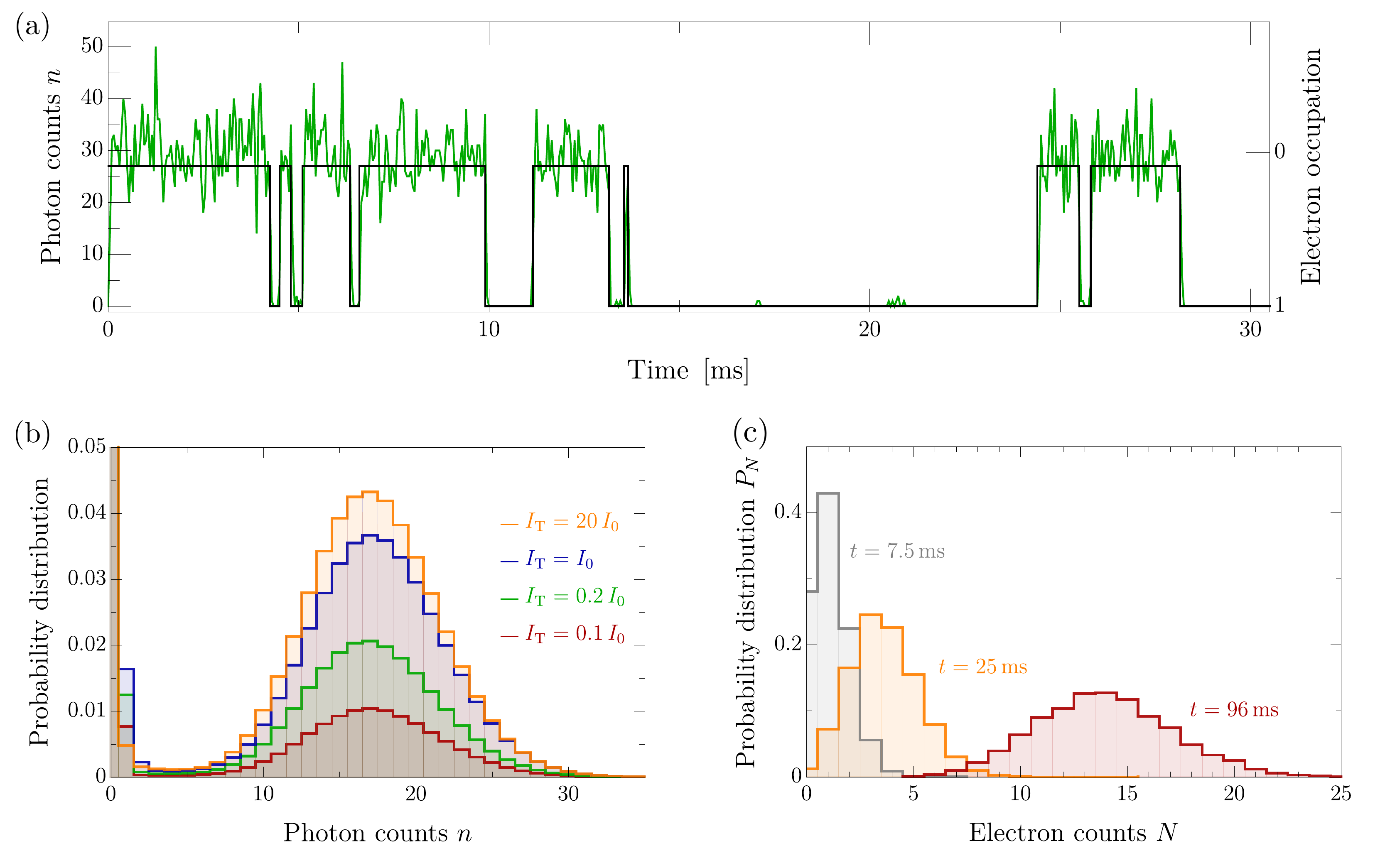}
  \caption{(a) Telegraph signal (black) of electron occupation derived from photon counts $n$ (green) with a threshold $n_\text{th}=4$.
  (b) Histogram of fluorescence photons with increasing trion laser intensity $I_\text{T}$. (c) Full counting statistics $P_N(t)$ for three different time intervals $t=7.5\,\text{ms}$ (gray), $t=25\,\text{ms}$ (orange), and $t=96\,\text{ms}$ (red). The binning time in (a)-(c) is $\Delta t=30\,\mu\text{s}$.
  }
  \label{fig:auger_histo}
\end{figure*}

\section{Charge fluctuations}\label{sec:auger_fcs}
The charge of the quantum dot is monitored in real time with the help of the exciton laser. 
It drives an exciton transition, $\ket{0}\leftrightarrow \ket{\text{X}}$, whenever the quantum dot is charge neutral, see Fig.~\ref{fig:auger_states}(a). 
The emitted fluorescence photons are detected by a single-photon detector as a function of time.
We count the number of photons $n$ that are emitted within the binning time of $\Delta t=30\,\mu\text{s}$.
An example of the resulting time trace is shown as a green line in Fig.~\ref{fig:auger_histo}(a).
Whenever the quantum dot is charge neutral, a large number of fluorescence photons are emitted.
Once an electron tunnels into the quantum dot, the exciton can no longer be excited, and the fluorescence signal drops to almost zero. Almost, because the fluorescence of the trion transition gives a small but (for our purpose) negligible contribution. 

In Fig.~\ref{fig:auger_histo}(b), we show the histograms of the detected fluorescence photons emitted from the quantum dot for four different trion laser intensities, $I_\text{T}=\lambda I_0$ with $\lambda=0.1, 0.2, 1$, and $20$, where $I_0=1.6 \times 10^{-5}\mu\text{W} /\mu \text{m}^2$ is some arbitrarily chosen reference intensity.
They all clearly display a bimodal distribution.
There is a broad peak around $n=17$ and a very narrow one at $n=0$.
The area covered by the broad peak indicates the probability to find the quantum dot empty, while the area of the narrow peak indicates the probability of an occupied quantum dot.
With increasing trion laser intensity $I_\text{T}\propto \Omega_\text{T}^2$, also the Rabi driving frequency $\Omega_\text{T}$ and, thus, the degree of trion saturation $n_\text{T}= \Omega_\text{T}^2/\gamma_{\uparrow\text{T}}^2$ increases, which leads to an increased Auger recombination rate $\Gamma_\text{A}=n_\text{T}\gamma_\text{A}$.
As a consequence, the probability to find the system in the empty (fluorescent) state increases\footnote{Note that there is also a small contribution of fluorescence photons from the trion excitation. However, the number of counts is much smaller and, thus, it merely contributes to the width of the peak at $n=0$ in Fig.~\ref{fig:auger_histo}(b).}. 
This is reflected by a larger area of the broad peak.
Since the photon count distributions are normalized to $1$, the increase of the broad peak is accompanied by a decrease of the height of the narrow peak [not visible in Fig.~\ref{fig:auger_histo}(b)].

To quantitatively study the charge fluctuations, we transform the binned photon stream (green line) into a binary signal of the electron occupation (black line) by introducing a threshold of $n_\text{th}=4$, see Fig.~\ref{fig:auger_histo}(a). 
Hence, for $n\le n_\text{th}$ the quantum dot is assumed to be occupied and for $n>n_\text{th}$ empty. 
The obtained random telegraph signal then contains the full information about the charge dynamics. 
In particular, we have access to the distribution of waiting times for an empty quantum dot to be filled and vice versa. We also obtain the full counting statistics describing the probability $P_N(t)$ that $N$ electrons have been ejected from the quantum dot in a time interval $t$, see Fig.~\ref{fig:auger_histo}(c).

We note here that in our measurement scheme the binning time $\Delta t$ can be chosen \textit{after} having measured the full time trace of all individual photon counts.
This opens the possibility to optimize a posteriori the choice of $\Delta t$ such that errors due to detector imperfections are suppressed~\cite{kerski_2023}. 
If the binning time is chosen too small then the narrow and the broad peaks in the photon counting statistics overlap which implies that false transitions are indicated by the detector although no tunneling event has occurred.
If, on the other hand, the binning time is chosen too large then fast sequences of tunneling-in and -out events may be overlooked by the detector.
While it is possible to model these sources of error theoretically~\cite{kleinherbers_2021_pushing}, it is more convenient to choose, if possible, $\Delta t$ such that they are negligible.
 
The dynamics of the system can be modelled by the rate equation
\begin{align}
\dot{\rho} ={\cal L}\rho=\begin{pmatrix}
-2\Gamma_\text{in} & 0& \Gamma_\text{A} & \\
\Gamma_\text{in} & -\Gamma_{\uparrow\downarrow} & \Gamma_\text{R}+ \Gamma_{\downarrow\uparrow} & \\
\Gamma_\text{in}  &  \Gamma_{\uparrow\downarrow}  & -\Gamma_\text{A}-\Gamma_\text{R}- \Gamma_{\downarrow\uparrow} &
\end{pmatrix} \rho,
\label{eq:auger_mastereq}
\end{align}
for the density matrix $\rho=\text{diag}\left(\rho_{0},\rho_{\downarrow},\rho_{\uparrow}\right)$ where $\rho_i$ is the probability for finding the quantum dot in state $i\in \{ 0,\downarrow,\uparrow \}$.
The density matrix $\rho_\text{st}$ for the stationary limit is determined by ${\cal L}\rho_\text{st}=0$.
While the Liouvillian $\cal L$ describes the full dynamics of the system, the processes of an electron tunneling into or out of the quantum dot are covered by the jump operators
\begin{align}
{\cal J}_\text{in}=\begin{pmatrix}
0& 0& 0 & \\
\Gamma_\text{in} & 0 &0 & \\
\Gamma_\text{in}  & 0 &0 & \\ 
\end{pmatrix} 
\quad \text{and}  \quad
{\cal J}_\text{out}=\begin{pmatrix}
0 & 0& \Gamma_\text{A} & \\
0 & 0& 0& \\
0 &  0 &0 & \\ 
\end{pmatrix},
\label{eq:auger_jumpop}
\end{align}
respectively.
The internal transitions within the singly-charged quantum dot, spin relaxation and spin-flip Raman scattering, are described by the remaining part of the Liouvillian, ${\cal L} - {\cal J}_\text{in} - {\cal J}_\text{out}$.

\subsection{Waiting-time distributions}
\begin{figure}[t]
  \includegraphics[width=.5\textwidth]{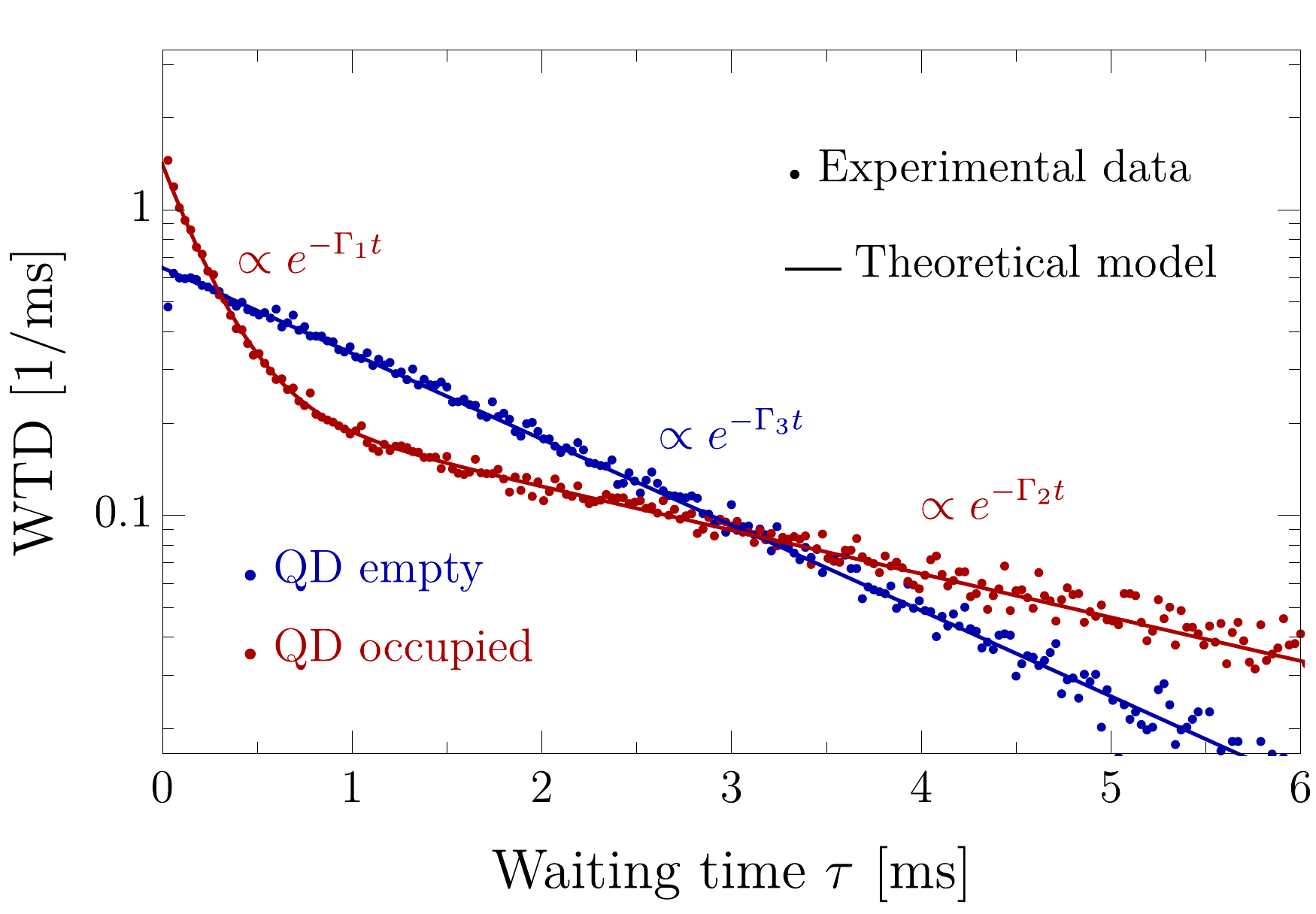}
  \caption{Waiting-time distributions $w_\text{emp}(\tau)$ (blue) and $w_\text{occ}(\tau)$ (red) showing how long the quantum dot (QD) is  empty and occupied, respectively. The parameters are $\Delta t=30\,\mu\text{s}$, $n_\text{th}=4$, and $I_\text{T}=I_0$.}
  \label{fig:wtd}
\end{figure}
One tool to characterize the measured charge fluctuations are waiting-time distributions~\cite{brandes_2008,albert_2012,rudge_2019,kleinherbers_2021_synchro}.
We define $w_\text{occ}(\tau)$ [with normalization $\int\limits_0^\infty d\tau w\, _\text{occ}(\tau) =1$] as the distribution of waiting times $\tau$ that describe how long the quantum dot is occupied (indicated by the absence of fluorescence) before it is emptied by an Auger recombination.
This is measured as the time distance between a tunneling-in event and an Auger process.
Similarly, $w_\text{emp}(\tau)$ describes the (normalized) distribution of how long the quantum dot stays empty (indicated by fluorescence) until an electron tunnels in, which corresponds to the waiting time between Auger and tunneling-in process. 

In some sense, waiting-time distributions resemble the idea of a pump-probe measurement.
Starting from a well-defined initial (charge) state, one probes as a function of time how this state decays into another (charge) state.
Technically, however, the time at which the system is prepared in the initial state is not given by some external stimulus but randomly chosen by the system itself.

The experimental result for the waiting times is displayed in Fig.~\ref{fig:wtd} for a given trion laser intensity $I_\text{T}=I_0$.
We find a bi-exponential behavior
\begin{align}
w_\text{occ}(\tau)= a \Gamma_1 e^{-\Gamma_1 \tau}+ (1-a) \Gamma_2 e^{-\Gamma_2 \tau},
\end{align} 
with rates $\Gamma_1=4.32\,\text{ms}^{-1}$ and $\Gamma_2=0.33\,\text{ms}^{-1}$ for the waiting-time distribution of an occupied quantum dot (red) and the weighting factor $a=0.27$.
For the empty quantum dot (blue), we get a mono-exponential decay
\begin{align}
w_\text{emp}(\tau)=\Gamma_3 e^{-\Gamma_3 \tau},
\end{align} 
with rate $\Gamma_3=0.65 
\,\text{ms}^{-1}$.
The mono-exponential decay indicates that there is only one transition rate for filling up an empty quantum dot, while the bi-exponential behavior has its origin in the internal spin dynamics that complicates the possibilities to empty the quantum dot:
In addition to an immediate ejection of an up-spin electron via an Auger process, a spin-down electron can leave the quantum dot in a two-step process by first relaxing to spin up and then ejection by the Auger effect.
As a result, the spin dynamics, either due to spin relaxation $\Gamma_{\uparrow\downarrow}$ and $\Gamma_{\downarrow\uparrow}$ or due to optical spin-flip Raman scattering, does not change the quantum-dot charge directly, but it does affect the charge dynamics indirectly.

Theoretically, the waiting-time distributions can be determined via~\cite{brandes_2008}
\begin{align}
w_\text{occ}(\tau) = \frac{\tr[{\cal J}_\text{out} e^{({\cal L}-{\cal J}_\text{in}-{\cal J}_\text{out})\tau} {\cal J}_\text{in} \rho_\text{st}]}{\tr[{\cal J}_\text{in} \rho_\text{st}]}, \label{eq:wtocc} \\
w_\text{emp}(\tau) = \frac{\tr[{\cal J}_\text{in} e^{({\cal L}-{\cal J}_\text{in}-{\cal J}_\text{out})\tau} {\cal J}_\text{out} \rho_\text{st}]}{\tr[{\cal J}_\text{out} \rho_\text{st}]} .\label{eq:wtemp}
\end{align}
After some algebra, we can express the relaxation rates $\Gamma_1$, $\Gamma_2$, and $\Gamma_3$ as well as the weighting factor $a$ of the (bi-)exponential decay in terms of the four independent transition rates $\Gamma_\text{in}$, $\Gamma_{\uparrow\downarrow}$, $\Gamma_\text{A}$, and $\Gamma_\text{R}$ that occur in the theoretical model.
We find the relations
\begin{align}
\Gamma_{1}-\Gamma_2&=\sqrt{\left(\Gamma_\text{A}-\Gamma_{\uparrow\downarrow}\right)^2+\Gamma_\text{P}\left[\Gamma_\text{P}+2\left(\Gamma_\text{A}+\Gamma_{\uparrow\downarrow}\right)\right]}, \nonumber \\
\Gamma_{1}+\Gamma_2&=\Gamma_\text{A}+\Gamma_{\uparrow\downarrow}+\Gamma_\text{P}, \nonumber \\
\Gamma_3&=2 \Gamma_\text{in}, \nonumber \\
a \Gamma_1+(1-a)\Gamma_2&= \frac{\Gamma_\text{A}}{2},
\end{align}
where we introduced as an abbreviation the total spin-pumping rate $\Gamma_\text{P}= {\Gamma}_\text{R}+e^{-\Delta_e/(k_\text{B} T)}{\Gamma}_{\uparrow\downarrow}$ as the sum of the spin-flip Raman and the inverse spin-relaxation rate.
This leads to the following transition rates
\begin{align}
&\Gamma_{\text{in}}=0.32\,\text{ms}^{-1}, \quad \Gamma_{\uparrow\downarrow}=0.50\,\text{ms}^{-1}, \nonumber \\  &\Gamma_\text{R}=1.02\,\text{ms}^{-1},  \quad \Gamma_\text{A}=2.82\,\text{ms}^{-1}.\label{eq:fitrates}
\end{align}
Hence, the Auger recombination has the fastest transition rate $\Gamma_\text{A}$, almost three times as large as the spin-flip Raman rate $\Gamma_\text{R}$. 
The latter is, in turn, larger than the spin-relaxation rate, $\Gamma_\text{R}>\Gamma_{\uparrow\downarrow}$, such that we can infer an inverted spin population.
In fact, the population of the high-energy spin state $\ket{\downarrow}$ has the highest probability
\begin{align}
\mel{0}{\rho_\text{st}}{0}&= 40.4\%, \nonumber \\ 
\mel{\downarrow}{\rho_\text{st}}{\downarrow}&= 50.3\% ,\nonumber \\  
\mel{\uparrow}{\rho_\text{st}}{\uparrow}&= 9.3\%.
\end{align}

It is important to notice that fitting the waiting-time distributions provides four numbers only.
For this to be sufficient, a number of assumptions have to be fulfilled since, in general, a model comprising three states can exhibit six rates in total. Here,
we assume equal rates $\gamma_{\uparrow 0}=\gamma_{\downarrow 0}$ for tunneling into the two different spin states of the quantum dot. We also make use of the fact that tunneling out of the dot is energetically forbidden. 

\subsection{Factorial cumulants}
\begin{figure}[t]
  \includegraphics[width=.5\textwidth]{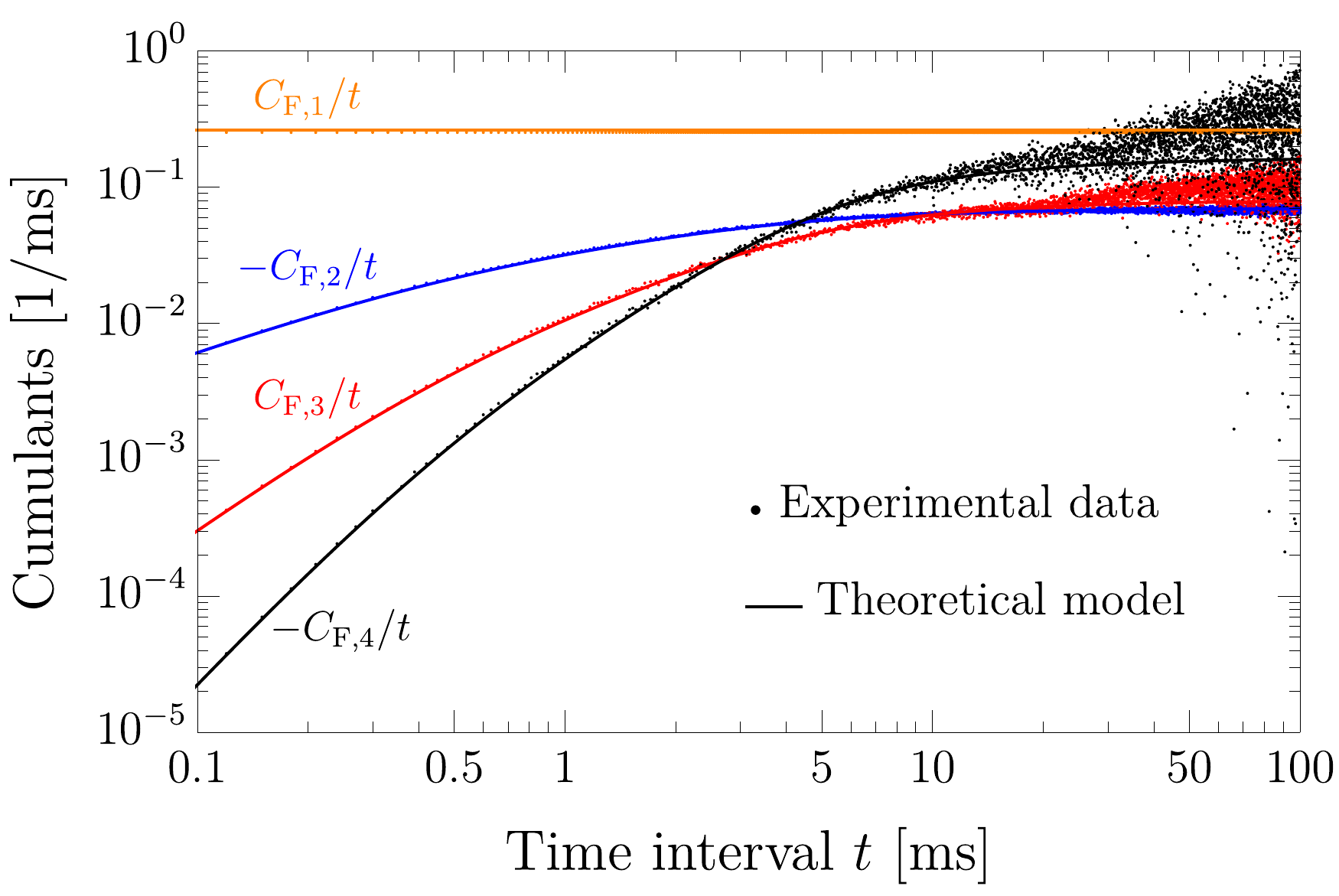}
  \caption{Factorial cumulants $C_{\text{F},m}$ as a function of time. Experimental data (dots) is compared with simulation (lines). The parameters are $\Delta t=30\,\mu\text{s}$, $n_\text{th}=4$, and $I_\text{T}=I_0$.
  }
  \label{fig:auger_faccum}
\end{figure}

In addition to waiting times, there is another convenient tool to address the charge-transfer statistics.
The full information of the latter is contained in the time-dependent probability distributions $P_N(t)$ to count $N$ Auger (i.e., ejecting-out) processes within a time interval of length $t$.
Alternatively, we could chose to count the tunneling-in events instead, which would yield, for the considered model, the same information.
Keeping track of the number $N$ of Auger recombinations, we extend our rate-equation description to
\begin{align}
\dot{\rho}_N = ({\cal L}- {\cal J}_\text{out})\rho_N + {\cal J}_\text{out}\rho_{N-1},
\end{align}
where $\rho_N$ is the density matrix with the constraint that $N$ Auger processes have been counted.

Performing a $z$-transform $\rho\z=\sum_N z^N \rho_N$ turns the rate equation into 
$\dot{\rho}\z = {\cal L}\z\rho\z$
with ${\cal L}\z = {\cal L} + (z-1){\cal J}_\text{out}$.
The formal solution 
$\rho\z = e^{{\cal L}\z t}\rho_\text{st}$
can be transformed back with the inverse $z$-transform to yield
\begin{align}
\rho_N(t)=\frac{1}{N!}\partial_z^N \left(e^{{\cal L}\z t}\rho_\text{st}\right)\vert_{z=0}.
\end{align}
Finally, we perform the trace over the quantum-dot states to arrive at the full counting statistics
\begin{align}\label{eq:pn}
P_N(t)=\tr [\rho_N(t)]=\frac{1}{N!}\partial_z^N \tr\left(e^{{\cal L}\z t}\rho_\text{st}\right)\vert_{z=0}.
\end{align}

For each interval length $t$, there is a distribution of the number $N$ of Auger events.
Such a discrete distribution is most conveniently characterized in terms of factorial cumulants $C_{\text{F},m}$~\cite{beenakker_counting_2001,flindt2009universal, kambly_2011,stegmann2015detection,stegmann2016short,kleinherbers2018revealing,koenig_2021,kleinherbers_2021_pushing}, which are the partners of factorial moments $M_{\text{F},m}=\langle N (N-1) \cdots (N-m+1)\rangle$.
The factorial cumulants can be obtained as derivatives
\begin{align}\label{eq:faccum}
C_{\text{F},m}=\partial_z^m {\cal S}(z,t)\vert_{z=1},
\end{align}
of the cumulant-generating function~\cite{stegmann2015detection,kleinherbers_2021_pushing}
\begin{align}\label{eq:facgen}
{\cal S}(z,t)=\ln \sum_Nz^NP_N(t)=\ln\tr\left(e^{{\cal L}\z t}\rho_\text{st}\right).
\end{align}
In the second step of Eq.~\eqref{eq:facgen}, we use Eq.~\eqref{eq:pn} and identify the Taylor expansion. 
(To obtain the factorial moments, one just needs to remove the logarithm from the generating function.) 
In the following, we use the first form of Eq.~\eqref{eq:facgen} to obtain the factorial cumulants from the experimental data, while the second form is more convenient for calculating them from the model.

We remark that for continuous stochastic variables, one often makes use of ordinary moments $M_m=\langle N^m \rangle$ and ordinary cumulants $C_m$, instead of factorial ones.
The ordinary cumulants $C_m$ can be obtained by replacing $z \rightarrow e^z$ in the generating function, and the ordinary moments $M_m$ by additionally removing the logarithm. 
In our case, however, the stochastic variable is discrete.
It is, then, more natural to use factorial cumulant \cite{koenig_2021}.
In fact, it is easy to show that for a Poissonian distribution, which defines the model of stochastically-independent discrete events as a reference, all factorial cumulants of order $m\ge 2$ vanish.
They have also the advantage that unwanted features of universal oscillations \cite{flindt2009universal} of the cumulants as a function of time and/or system parameters are avoided.
Furthermore, factorial cumulants are better suited to identify correlations between the individual counted events \cite{beenakker_counting_2001,kambly_2011,stegmann2015detection,stegmann2016short,kleinherbers2018revealing}. 
The most important reason, however, to use factorial cumulants instead of ordinary ones is that they are intrinsically resilient to imperfections of the detector caused by a finite bandwidth or false noise-induced counting events \cite{kleinherbers_2021_pushing}.

In Fig.~\ref{fig:auger_faccum}, we show the first four factorial cumulants as a function of the time interval length $t$ for our model, where the rates were taken from Eq.~\eqref{eq:fitrates} of the waiting-time analysis.
We find for all times $t$ a perfect agreement between the experimental data (dots) and theoretical curves (solid lines).
The data for the higher-order factorial cumulants (not shown) are more noisy but we again find good agreement between theory and experiment.

We would like to point out that this excellent agreement in the factorial-cumulants analysis was achieved, even though the necessary parameters were taken from a complementary data evaluation, i.e. the waiting time distribution. While the latter  only takes time intervals with exactly one tunneling-in and one Auger-recombination process into account, the former covers arbitrarily long time intervals. This not only provides a valuable benchmark for the statistical evaluation by factorial cumulants, a technique that has recently gained some popularity. It also shows that the model and its underlying assumptions are well suited to describe the spin and charge dynamics in the investigated system, as sketched in Fig.~\ref{fig:auger_states}. 

\subsection{Beyond the detector resolution}

\begin{figure}[t]
  \includegraphics[width=.5\textwidth]{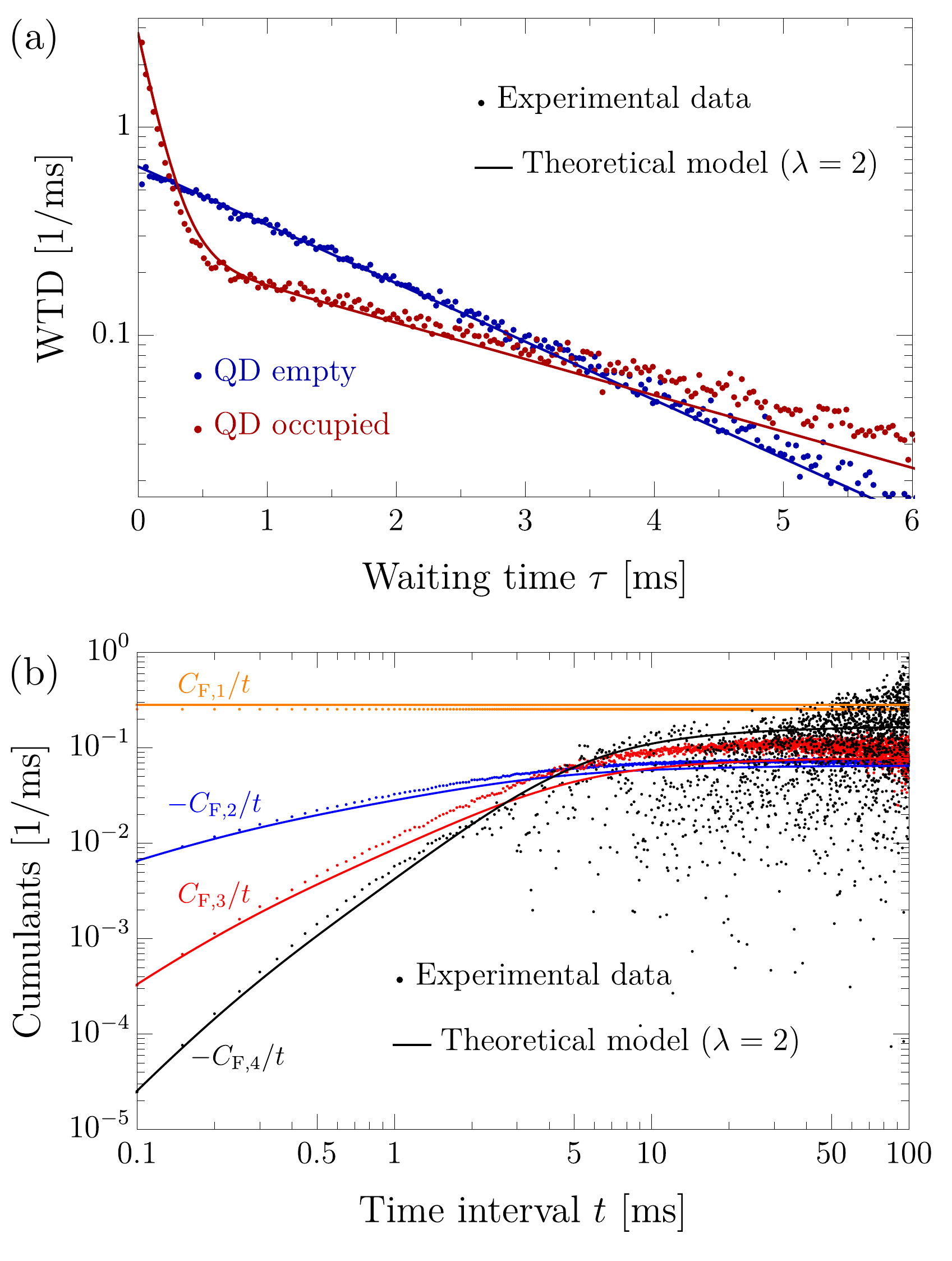}
  \caption{Moderate increase of the trion laser intensity with $\lambda=2$. (a) Waiting-time distribution and (b) factorial cumulants $C_{\text{F},m}$. Experimental data (dots) is compared with our theoretical model (lines).  For (a), we use $\Delta t=30\,\mu\text{s}$ and $n_\text{th}=4$ and for (b), we use $\Delta t=50\,\mu\text{s}$ and $n_\text{th}=10$.}
  \label{fig:facs_wtd_moder}
\end{figure}
Both the Auger and the spin-flip Raman rate depend on the trion occupation probability $n_\text{T}$.
The latter, in turn, depends on the trion laser intensity.
We repeat the experiment with different increased laser intensities, $I_\text{T}=I_0\rightarrow \lambda I_0$ and $\lambda>1$.
Since $I_\text{T}\propto \Omega_\text{T}^2$ and also $n_\text{T}\propto\Omega_\text{T}^2$ for $\gamma_{\uparrow\text{T}}\gg \Omega_\text{T}$, we expect from Eq.~\eqref{eq:auger_effrates} that the rates of the optically induced Auger effect and the spin-flip Raman scattering increase by the same factor as the laser intensity
\begin{align}\label{eq:auger_newrates}
\Gamma_\text{R}\rightarrow \lambda \Gamma_\text{R}\quad  \text{and} \quad \Gamma_\text{A}\rightarrow \lambda \Gamma_\text{A}.
\end{align}
All other rates should remain unchanged.

\subsubsection{Moderate laser-intensity increase}

First, we increase the trion laser intensity by a factor of $\lambda=2$ only.
In this case, we only change the Auger and spin-flip Raman rates in the theoretical modelling. 
The derived occupation probabilities are now
\begin{align}
\mel{0}{\rho_\text{st}}{0}&= 43.6\%, \nonumber \\ 
\mel{\downarrow}{\rho_\text{st}}{\downarrow}&= 51.4\% ,\nonumber \\  \mel{\uparrow}{\rho_\text{st}}{\uparrow}&= 5\%.
\end{align}

For the experimental determination of waiting times we still choose $\Delta t=30 \, \mu \text{s}$ and  $n_\text{th}=4$. However, for the full counting statistics, we increase the photon binning time slightly and adjust the threshold accordingly.
If we kept a binning time of $\Delta t = 30\,\mu\text{s}$ as before then the overlap of the narrow and broad peak in the photon counting statistics, see Fig.~\ref{fig:auger_histo}(b), would start to play a role.
This increases the noise on the photon-count signal, leading to false counts.
To avoid these false counts, we increase the binning time to $\Delta t = 50\,\mu\text{s}$.
This shifts the maximum of the broad peak of the photon counting statistics to $\sim 30$.
For the threshold to discriminate the empty from the occupied quantum dot, we now choose $n_\text{th}=10$
in order to avoid too many false (noise-induced) events.

In Fig.~\ref{fig:facs_wtd_moder} we show the comparison between theory and experiment for both the waiting-time distributions and the factorial cumulants. 
We find fairly good but not perfect agreement. Deviations occur from systematic detection errors of missing counts as discussed in the following.

\begin{figure*}[t]
  \includegraphics[width=.95\textwidth]{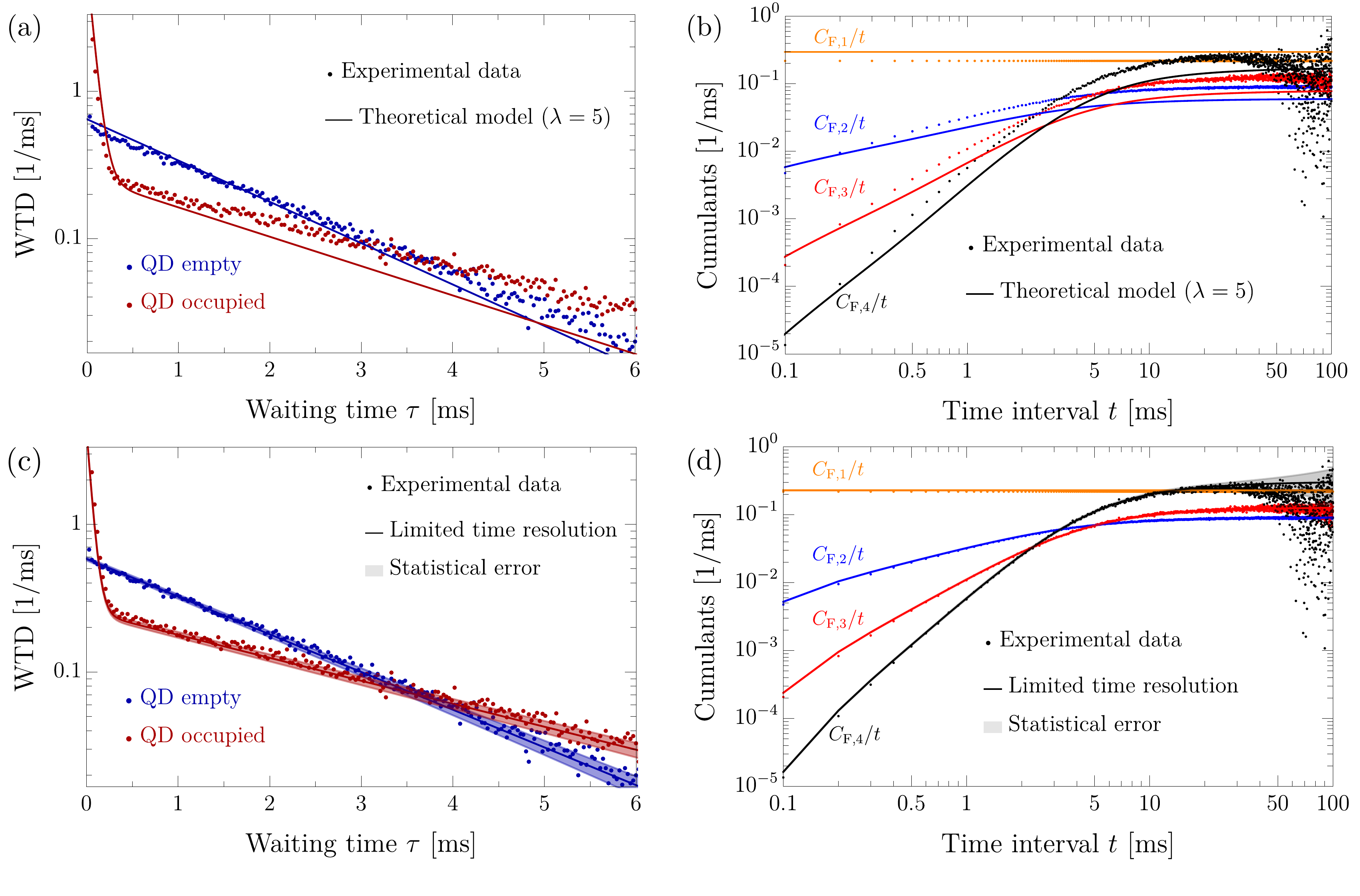}
  \caption{Strong increase of the trion laser intensity with $\lambda=5$. 
  (a) and (c) show the waiting-time distribution (for $\Delta t=30\,\mu\text{s}$ and $n_\text{th}=4$), (b) and (d) the factorial cumulants $C_{\text{F},m}$ (for $\Delta t=100\,\mu\text{s}$ and $n_\text{th}=30$). 
  Experimental data (dots) is compared with our theoretical model (lines).
  The theoretical model used in (a) and (b) is continuous assuming a perfect detector, the one in (c) and (d) is discrete including systematic errors due to a limited time resolution (solid lines connect discrete values at times $n \Delta t$ with $n= 0,1,\ldots$) and statistical errors due to a finite measurement time (shaded background).
  We find very good agreement between experiment and theory only when taking the limited time resolution into account in the theoretical description.
}
  \label{fig:facs_wtd_high}
\end{figure*}
\subsubsection{Strong laser-intensity increase}

Increasing the trion laser intensity by a factor of $\lambda=5$ leads to the occupation probabilities
\begin{align}
\mel{0}{\rho_\text{st}}{0}&=45.8\%, \nonumber \\ 
\mel{\downarrow}{\rho_\text{st}}{\downarrow}&=52.1\% ,  \nonumber \\\mel{\uparrow}{\rho_\text{st}}{\uparrow}&=2.1\% ,
\end{align}
i.e., the spin-up state is almost entirely depleted. More importantly, a strong inversion of the spin population (factor 25) is found.
An even further increase of the laser intensity to $\lambda=20$ [corresponding to the orange curve in Fig.~\ref{fig:auger_histo}(b)] changes these probabilities only slightly.

For $\lambda=5$, we again choose for the waiting times $\Delta t=30 \, \mu \text{s}$ and  $n_\text{th}=4$, while for the full counting statistics we increase the photon binning time to $\Delta t=100 \, \mu \text{s}$ and the threshold to $n_\text{th}=30$.
The resulting waiting-time distributions and factorial cumulants $C_{\text{F},m}$ as a function of time $t$ are shown in  Fig.~\ref{fig:facs_wtd_high}(a) and (b), respectively.
We find that the experimental data (dots) strongly deviates from the theoretical curves (lines), which were obtained by the simple replacement Eq.~\eqref{eq:auger_newrates}.
Here, the Auger rate has become so large that it is comparable to the detector bandwidth
\begin{align}
\Gamma_\text{A} \Delta t \sim 1 ,
\end{align}
so that the time resolution of the detection scheme is now too limited to keep track of every Auger-recombination process and errors stemming from missing counts can no longer be ignored.
Since it is impossible to find a binning time that prevents both false and missing counts,
we proceed by keeping the long binning time but include the systematic error due to the limited time resolution into our theory for both the waiting-time distribution and the factorial cumulants. 

Let us start with the full counting statistics of the Auger recombination processes.
We follow Ref.~\cite{kleinherbers_2021_pushing} and replace the continuous time evolution $e^{{\cal L}\z t}$ by a $t/\Delta t$-fold iteration of the \textit{finite}-time propagation 
\begin{align}
{\Pi\z}= 
	\left(\begin{array}{ccc}
	1& z& z  \\  
	1& 1& 1 \\
	1& 1& 1\\
	\end{array}\right) \circ e^{{\cal L}_1\Delta t}
\end{align}
for a time step $\Delta t$.
Here, $\circ$ denotes the Hadamard product, i.e., an element-wise multiplication between the two matrices. 
By construction, we have $\Pi\z\neq e^{{\cal L}\z\Delta t}$, which emphasizes that the counting variable $z$ is only introduced \textit{after} coarse graining the time evolution. 
Thereby, we ensure that during each time step $\Delta t$ at most one Auger charge transfer is counted which effectively simulates missing events on smaller time scales. 
With this coarse-grained time evolution, Eq.~\eqref{eq:facgen} has to be replaced by the cumulant-generating function
\begin{align}
{\cal S}(z,t)=\ln\tr\left({\Pi\z}^{t/\Delta t} \rho_\text{st}\right).
\end{align}
Now we can derive, by again using Eq.~\eqref{eq:faccum}, the cumulants $C_{\text{F},m}$ which include the systematic error due to a limited time resolution. 
The comparison between experimental data (dots) and improved theory (lines) is shown in Fig.~\ref{fig:facs_wtd_high}(d).
Perfect agreement is found.
We can even theoretically model the statistical error $\Delta C_{\text{F},m}$ due to a finite measurement time $\TX$ of the telegraph signal, indicated by a shaded background. 
Its derivation, based on Ref.~\cite{kleinherbers_2021_pushing} is presented in the Appendix~\ref{sec:stat}. 

For the waiting-time distributions, we proceed in a similar manner.
In this case, we need to introduce the coarse-grained jump operators  
\begin{align}
{\Pi}_\text{in}= 
	\left(\begin{array}{ccc}
	0& 0& 0  \\  
	1& 0& 0 \\
	1& 0& 0\\
	\end{array}\right) \circ e^{{\cal L}_1\Delta t}
\end{align}
and
\begin{align}
{\Pi}_\text{out}= 
	\left(\begin{array}{ccc}
	0& 1& 1  \\  
	0& 0& 0 \\
	0& 0& 0\\
	\end{array}\right) \circ e^{{\cal L}_1\Delta t}.
\end{align}
By defining the full propagator $\Pi=e^{{\cal L}\Delta t}$ for a time interval $\Delta t$, we can replace the continuous time evolution $e^{({\cal L}-{\cal J}_\text{in}-{\cal J}_\text{out})\tau}$
in the absence of a charge transfer, see Eq.~\eqref{eq:wtocc}-\eqref{eq:wtemp}, by a $\tau/\Delta t$-fold iteration of the finite-time propagation $\Pi{-}\Pi_\text{in}{-}\Pi_\text{out}$. By also replacing the jump operators ${\cal J}_\text{in/out}$ by $\Pi_\text{in/out}$, this yields the waiting-time distributions\begin{align}
w_\text{occ}(\tau) =\frac{1}{\Delta t} \frac{\tr[{\Pi}_\text{out} \left(\Pi{-}\Pi_\text{in}{-}\Pi_\text{out}\right)^{\tau/\Delta t} {\Pi}_\text{in} \rho_\text{st}]}{\tr[{\Pi}_\text{in} \rho_\text{st}]}, \\
w_\text{emp}(\tau) =\frac{1}{\Delta t} \frac{\tr[{\Pi}_\text{in} \left(\Pi{-}\Pi_\text{in}{-}\Pi_\text{out}\right)^{\tau/\Delta t} {\Pi}_\text{out} \rho_\text{st}]}{\tr[{\Pi}_\text{out} \rho_\text{st}]},
\end{align}
which, summed over the discrete waiting times $\tau=n \Delta t$ with $n\in 1,2,\ldots$, fulfill the normalization $\sum_n \Delta t \,w_\text{occ/emp}(n\Delta t)=1$.
Once the finite time resolution is taken into account, the agreement between theory and experiment is very good, as shown in Fig.~\ref{fig:facs_wtd_high}(c). Again, we  theoretically model the statistical error $\Delta w_{\text{occ/emp}}$ due to a finite measurement time of the telegraph signal, indicated by a shaded background. 
Its derivation is presented in the Appendix~\ref{sec:stat_wt}. 

\section{Conclusions}\label{sec:auger_conclusion}

We studied the charge and spin dynamics of a self-assembled InAs quantum dot coupled to a laser field that drives an optical trion transition. 
We measured the charge fluctuations in form of a random telegraph signal by an optical readout scheme. 
By analyzing the charge fluctuations using waiting-time distributions and full counting statistics, we were able to fully describe the dynamics of the quantum dot. 
We determined not only the charge dynamics due to the Auger process and electron tunneling, but we also revealed the interal spin dynamics due to spin relaxation and optical spin pumping via spin-flip Raman processes. 
We showed that the optical spin pumping leads to a significant population inversion of the spin states and determined the electron tunneling rate, the Auger rate, the spin-relaxation rate and the rate of optical spin pumping in one go. Remarkably, these dynamic properties could be extracted from a steady-state experiment, where neither the electrical nor the optical parameters needed to be pulsed or modulated in time. Our evaluation of the optical random telegraph signal is therefore a nice illustration of Rolf Landauer's famous statement that ``the noise is the signal''~\cite{landauer_1998}. 
Finally, we have shown that both complementary statistical frameworks used for the data analysis (i.e. waiting-time distributions and  factorial cumulants) are in excellent agreement with each other and with the experimental data. Even in situations, where the experimental data is distorted by a limited time resolution, it can accurately be modeled by the employed statistical methods.  

\begin{acknowledgments}
This work was financially supported by the Deutsche Forschungsgemeinschaft (DFG, German Research Foundation) under Project-ID 278162697 -- SFB 1242 and the MERCUR project Ko-2022-0013.
\end{acknowledgments}

\appendix
\section{Statistical error of factorial cumulants}~\label{sec:stat}
The statistical error of the full counting statistics stemming from the finite time $\TX$ of the telegraph signal lead to a variance of the factorial cumulants (in Fig.~\ref{fig:facs_wtd_high}(d) we have $\TX=335\,\text{s}$).
As shown in the Supplemental Material of Ref.~\cite{kleinherbers_2021_pushing}, it can be expressed as
\begin{align}\label{eq:DeltaFC}
\left(\Delta C_{\text{F},m}\right)^2=\frac{t}{\TX} \sum_{k,k^\prime} \frac{\partial C_{\text{F},m}}{\partial M_k}\frac{\partial C_{\text{F},m}}{\partial M_{k^\prime}} \left(M_{k+k^\prime}-M_k M_{k^\prime}\right),
\end{align}
in terms of the ordinary moments $M_m=\ev{N^m}$. 
To evaluate the right-hand side of Eq.~\eqref{eq:DeltaFC}, we need the functional relation $C_{\text{F},m}=C_{\text{F},m}(M_1,\ldots,M_m)$ of the factorial cumulants in terms of the ordinary moments.
This is achieved in two steps.
First, we use the recursion formula~\cite{smith_1995}
\begin{align}\label{eq:intro_fcs_ordcumofm}
C_{\text{F},m}=M_{\text{F},m} - \sum_{k=1}^{m-1}\binom{m{-}1}{k{-}1} C_{\text{F},k} M_{\text{F},m-k} 
\end{align}
relating factorial cumulants $C_{\text{F},m}$ to factorial moments $M_{\text{F},m}=\ev{N(N-1)\ldots(N-m+1)}$.
Second, we employ the linear relation between factorial and ordinary moments~\cite{johnson_univariate_2005}
\begin{align}\label{eq:intro_fcs_ordtofac}
M_{\text{F},m}=\sum_{k=1}^m s_1(m,k) M_k,
\end{align} 
with $s_1(m,k)$ being the Stirling numbers of the first kind. 
After performing the derivatives in Eq.~\eqref{eq:DeltaFC}, we reexpress the resulting function of ordinary moments in terms of factorial cumulants.
As a final result, we find 
\begin{align}
\left(\Delta C_{\text{F},1}\right)^2&={\frac{t}{\TX}} \left(C_{\text{F},1}+C_{\text{F},2}\right)={\frac{t}{\TX}} \left(\ev{N^2}-\ev{N}^2\right),\\
\left(\Delta C_{\text{F},2}\right)^2&={\frac{t}{\TX}}\left[2\left(C_{\text{F},1}+C_{\text{F},2}\right)^2+2C_{\text{F},2}+4C_{\text{F},3}+C_{\text{F},4}\right],
\end{align}
for the first two factorial cumulants and similar analytic expressions for $m\ge 3$, which are, however, too long to be presented here. 
The first line is the known result from the central limit theorem describing how the sample mean (formed from $\TX/t$ samples)  approaches the expectation value. 
Finally, we remark that in the long-time limit the variance of all factorial cumulants acquires a simple form again.
One finds the very compact expression~\cite{kleinherbers_2021_pushing} 
\begin{align}
\left(\Delta C_{\text{F},m}\right)^2\simeq\frac{m!\, t}{\TX}\left({\ev{N^2}}{-}{\ev{N}}^2\right)^m,
\end{align}
which is proportional to $t^{m+1}$.
This nicely shows how an increased order $m$ and longer time $t$ give rise to an increased statistical error.

\section{Statistical error of the waiting-time distribution}~\label{sec:stat_wt}
Due to the finite measurement time $\TX$, there is only a finite number $K$ of measured waiting times (in Fig.~\ref{fig:facs_wtd_high}(c) we have $K=85\,343$) which are all of the form $\tau=n\Delta t$ with $n\in 1,2,\ldots$. 
To experimentally determine the probability $w_\text{occ}(m \Delta t) \Delta t$ that a given (occupied) waiting time $\tau=m\Delta t$ occurs, we calculate relative frequencies by formally performing the sample mean $\bar{X}=\sum_n X_{n}/K$ over the quantity $X_n=\delta_{m,n}$. From the central limit theorem, we know that the sample mean $\bar{X}$ deviates from the expectation value $\ev{X}$ with
\begin{align}
    \left(\bar X-\ev{X}\right)^2\simeq\frac{1}{K}\left(\ev{X^2}-\ev{X}^2\right),
\end{align}
where the expectation value is given by the exact probability $\ev{X}=w_\text{occ} \Delta t$ and the variance is given by ${\ev{X^2}}{-}\ev{X}^2=w_\text{occ}\Delta t\left(1- w_\text{occ}\Delta t\right)$ with $w_\text{occ}=w_\text{occ}(m\Delta t)$.
Thus, the statistical error of the waiting time distribution can be estimated as 
\begin{align}
&\Delta w_\text{occ}(\tau)=\frac{1}{\Delta t \sqrt{K}}\sqrt{ w_\text{occ}(\tau)\Delta t\left[1- w_\text{occ}(\tau)\Delta t\right]},\\
&\Delta w_\text{emp}(\tau)=\frac{1}{\Delta t \sqrt{K}}\sqrt{w_\text{emp}(\tau)\Delta t\left[1- w_\text{emp}(\tau)\Delta t\right]}.
\end{align}
In Fig.~\ref{fig:facs_wtd_high}(c), we see that the analytical estimates of the statistical error correctly describe the observed fluctuations in the experimental data.

\bibliography{references}

\begin{thebibliography}{60}%
\makeatletter
\providecommand \@ifxundefined [1]{%
 \@ifx{#1\undefined}
}%
\providecommand \@ifnum [1]{%
 \ifnum #1\expandafter \@firstoftwo
 \else \expandafter \@secondoftwo
 \fi
}%
\providecommand \@ifx [1]{%
 \ifx #1\expandafter \@firstoftwo
 \else \expandafter \@secondoftwo
 \fi
}%
\providecommand \natexlab [1]{#1}%
\providecommand \enquote  [1]{``#1''}%
\providecommand \bibnamefont  [1]{#1}%
\providecommand \bibfnamefont [1]{#1}%
\providecommand \citenamefont [1]{#1}%
\providecommand \href@noop [0]{\@secondoftwo}%
\providecommand \href [0]{\begingroup \@sanitize@url \@href}%
\providecommand \@href[1]{\@@startlink{#1}\@@href}%
\providecommand \@@href[1]{\endgroup#1\@@endlink}%
\providecommand \@sanitize@url [0]{\catcode `\\12\catcode `\$12\catcode
  `\&12\catcode `\#12\catcode `\^12\catcode `\_12\catcode `\%12\relax}%
\providecommand \@@startlink[1]{}%
\providecommand \@@endlink[0]{}%
\providecommand \url  [0]{\begingroup\@sanitize@url \@url }%
\providecommand \@url [1]{\endgroup\@href {#1}{\urlprefix }}%
\providecommand \urlprefix  [0]{URL }%
\providecommand \Eprint [0]{\href }%
\providecommand \doibase [0]{https://doi.org/}%
\providecommand \selectlanguage [0]{\@gobble}%
\providecommand \bibinfo  [0]{\@secondoftwo}%
\providecommand \bibfield  [0]{\@secondoftwo}%
\providecommand \translation [1]{[#1]}%
\providecommand \BibitemOpen [0]{}%
\providecommand \bibitemStop [0]{}%
\providecommand \bibitemNoStop [0]{.\EOS\space}%
\providecommand \EOS [0]{\spacefactor3000\relax}%
\providecommand \BibitemShut  [1]{\csname bibitem#1\endcsname}%
\let\auto@bib@innerbib\@empty
\bibitem [{\citenamefont {Loss}\ and\ \citenamefont
  {DiVincenzo}(1998)}]{loss_1998}%
  \BibitemOpen
  \bibfield  {author} {\bibinfo {author} {\bibfnamefont {D.}~\bibnamefont
  {Loss}}\ and\ \bibinfo {author} {\bibfnamefont {D.~P.}\ \bibnamefont
  {DiVincenzo}},\ }\bibfield  {title} {\bibinfo {title} {Quantum computation
  with quantum dots},\ }\href {https://doi.org/10.1103/PhysRevA.57.120}
  {\bibfield  {journal} {\bibinfo  {journal} {Phys. Rev. A}\ }\textbf {\bibinfo
  {volume} {57}},\ \bibinfo {pages} {120} (\bibinfo {year} {1998})}\BibitemShut
  {NoStop}%
\bibitem [{\citenamefont {Elzerman}\ \emph {et~al.}(2004)\citenamefont
  {Elzerman}, \citenamefont {Hanson}, \citenamefont {Willems~van Beveren},
  \citenamefont {Witkamp}, \citenamefont {Vandersypen},\ and\ \citenamefont
  {Kouwenhoven}}]{elzerman_2004}%
  \BibitemOpen
  \bibfield  {author} {\bibinfo {author} {\bibfnamefont {J.~M.}\ \bibnamefont
  {Elzerman}}, \bibinfo {author} {\bibfnamefont {R.}~\bibnamefont {Hanson}},
  \bibinfo {author} {\bibfnamefont {L.~H.}\ \bibnamefont {Willems~van
  Beveren}}, \bibinfo {author} {\bibfnamefont {B.}~\bibnamefont {Witkamp}},
  \bibinfo {author} {\bibfnamefont {L.~M.~K.}\ \bibnamefont {Vandersypen}},\
  and\ \bibinfo {author} {\bibfnamefont {L.~P.}\ \bibnamefont {Kouwenhoven}},\
  }\bibfield  {title} {\bibinfo {title} {Single-shot read-out of an individual
  electron spin in a quantum dot},\ }\href
  {https://doi.org/10.1038/nature02693} {\bibfield  {journal} {\bibinfo
  {journal} {Nature}\ }\textbf {\bibinfo {volume} {430}},\ \bibinfo {pages}
  {431} (\bibinfo {year} {2004})}\BibitemShut {NoStop}%
\bibitem [{\citenamefont {Amasha}\ \emph {et~al.}(2008)\citenamefont {Amasha},
  \citenamefont {MacLean}, \citenamefont {Radu}, \citenamefont {Zumb\"uhl},
  \citenamefont {Kastner}, \citenamefont {Hanson},\ and\ \citenamefont
  {Gossard}}]{amasha_2008}%
  \BibitemOpen
  \bibfield  {author} {\bibinfo {author} {\bibfnamefont {S.}~\bibnamefont
  {Amasha}}, \bibinfo {author} {\bibfnamefont {K.}~\bibnamefont {MacLean}},
  \bibinfo {author} {\bibfnamefont {I.~P.}\ \bibnamefont {Radu}}, \bibinfo
  {author} {\bibfnamefont {D.~M.}\ \bibnamefont {Zumb\"uhl}}, \bibinfo {author}
  {\bibfnamefont {M.~A.}\ \bibnamefont {Kastner}}, \bibinfo {author}
  {\bibfnamefont {M.~P.}\ \bibnamefont {Hanson}},\ and\ \bibinfo {author}
  {\bibfnamefont {A.~C.}\ \bibnamefont {Gossard}},\ }\bibfield  {title}
  {\bibinfo {title} {Electrical {C}ontrol of {S}pin {R}elaxation in a {Q}uantum
  {D}ot},\ }\href {https://doi.org/10.1103/PhysRevLett.100.046803} {\bibfield
  {journal} {\bibinfo  {journal} {Phys. Rev. Lett.}\ }\textbf {\bibinfo
  {volume} {100}},\ \bibinfo {pages} {046803} (\bibinfo {year}
  {2008})}\BibitemShut {NoStop}%
\bibitem [{\citenamefont {Ladd}\ \emph {et~al.}(2010)\citenamefont {Ladd},
  \citenamefont {Jelezko}, \citenamefont {Laflamme}, \citenamefont {Nakamura},
  \citenamefont {Monroe},\ and\ \citenamefont {O'Brien}}]{ladd_2010}%
  \BibitemOpen
  \bibfield  {author} {\bibinfo {author} {\bibfnamefont {T.~D.}\ \bibnamefont
  {Ladd}}, \bibinfo {author} {\bibfnamefont {F.}~\bibnamefont {Jelezko}},
  \bibinfo {author} {\bibfnamefont {R.}~\bibnamefont {Laflamme}}, \bibinfo
  {author} {\bibfnamefont {Y.}~\bibnamefont {Nakamura}}, \bibinfo {author}
  {\bibfnamefont {C.}~\bibnamefont {Monroe}},\ and\ \bibinfo {author}
  {\bibfnamefont {J.~L.}\ \bibnamefont {O'Brien}},\ }\bibfield  {title}
  {\bibinfo {title} {Quantum computers},\ }\href
  {https://doi.org/10.1038/nature08812} {\bibfield  {journal} {\bibinfo
  {journal} {Nature}\ }\textbf {\bibinfo {volume} {464}},\ \bibinfo {pages}
  {45} (\bibinfo {year} {2010})}\BibitemShut {NoStop}%
\bibitem [{\citenamefont {Petta}\ \emph {et~al.}(2005)\citenamefont {Petta},
  \citenamefont {Johnson}, \citenamefont {Taylor}, \citenamefont {Laird},
  \citenamefont {Yacoby}, \citenamefont {Lukin}, \citenamefont {Marcus},
  \citenamefont {Hanson},\ and\ \citenamefont {Gossard}}]{petta_2005}%
  \BibitemOpen
  \bibfield  {author} {\bibinfo {author} {\bibfnamefont {J.~R.}\ \bibnamefont
  {Petta}}, \bibinfo {author} {\bibfnamefont {A.~C.}\ \bibnamefont {Johnson}},
  \bibinfo {author} {\bibfnamefont {J.~M.}\ \bibnamefont {Taylor}}, \bibinfo
  {author} {\bibfnamefont {E.~A.}\ \bibnamefont {Laird}}, \bibinfo {author}
  {\bibfnamefont {A.}~\bibnamefont {Yacoby}}, \bibinfo {author} {\bibfnamefont
  {M.~D.}\ \bibnamefont {Lukin}}, \bibinfo {author} {\bibfnamefont {C.~M.}\
  \bibnamefont {Marcus}}, \bibinfo {author} {\bibfnamefont {M.~P.}\
  \bibnamefont {Hanson}},\ and\ \bibinfo {author} {\bibfnamefont {A.~C.}\
  \bibnamefont {Gossard}},\ }\bibfield  {title} {\bibinfo {title} {{Coherent
  Manipulation of Coupled Electron Spins in Semiconductor Quantum Dots}},\
  }\href {https://doi.org/10.1126/science.1116955} {\bibfield  {journal}
  {\bibinfo  {journal} {Science}\ }\textbf {\bibinfo {volume} {309}},\ \bibinfo
  {pages} {2180} (\bibinfo {year} {2005})}\BibitemShut {NoStop}%
\bibitem [{\citenamefont {Imamo{\=g}lu}\ \emph {et~al.}(1999)\citenamefont
  {Imamo{\=g}lu}, \citenamefont {Awschalom}, \citenamefont {Burkard},
  \citenamefont {DiVincenzo}, \citenamefont {Loss}, \citenamefont {Sherwin},\
  and\ \citenamefont {Small}}]{imamoglu_1999}%
  \BibitemOpen
  \bibfield  {author} {\bibinfo {author} {\bibfnamefont {A.}~\bibnamefont
  {Imamo{\=g}lu}}, \bibinfo {author} {\bibfnamefont {D.~D.}\ \bibnamefont
  {Awschalom}}, \bibinfo {author} {\bibfnamefont {G.}~\bibnamefont {Burkard}},
  \bibinfo {author} {\bibfnamefont {D.~P.}\ \bibnamefont {DiVincenzo}},
  \bibinfo {author} {\bibfnamefont {D.}~\bibnamefont {Loss}}, \bibinfo {author}
  {\bibfnamefont {M.}~\bibnamefont {Sherwin}},\ and\ \bibinfo {author}
  {\bibfnamefont {A.}~\bibnamefont {Small}},\ }\bibfield  {title} {\bibinfo
  {title} {Quantum {I}nformation {P}rocessing {U}sing {Q}uantum {D}ot {S}pins
  and {C}avity {QED}},\ }\href {https://doi.org/10.1103/PhysRevLett.83.4204}
  {\bibfield  {journal} {\bibinfo  {journal} {Phys. Rev. Lett.}\ }\textbf
  {\bibinfo {volume} {83}},\ \bibinfo {pages} {4204} (\bibinfo {year}
  {1999})}\BibitemShut {NoStop}%
\bibitem [{\citenamefont {Xu}\ \emph {et~al.}(2007)\citenamefont {Xu},
  \citenamefont {Wu}, \citenamefont {Sun}, \citenamefont {Huang}, \citenamefont
  {Cheng}, \citenamefont {Steel}, \citenamefont {Bracker}, \citenamefont
  {Gammon}, \citenamefont {Emary},\ and\ \citenamefont {Sham}}]{xu_2007}%
  \BibitemOpen
  \bibfield  {author} {\bibinfo {author} {\bibfnamefont {X.}~\bibnamefont
  {Xu}}, \bibinfo {author} {\bibfnamefont {Y.}~\bibnamefont {Wu}}, \bibinfo
  {author} {\bibfnamefont {B.}~\bibnamefont {Sun}}, \bibinfo {author}
  {\bibfnamefont {Q.}~\bibnamefont {Huang}}, \bibinfo {author} {\bibfnamefont
  {J.}~\bibnamefont {Cheng}}, \bibinfo {author} {\bibfnamefont {D.~G.}\
  \bibnamefont {Steel}}, \bibinfo {author} {\bibfnamefont {A.~S.}\ \bibnamefont
  {Bracker}}, \bibinfo {author} {\bibfnamefont {D.}~\bibnamefont {Gammon}},
  \bibinfo {author} {\bibfnamefont {C.}~\bibnamefont {Emary}},\ and\ \bibinfo
  {author} {\bibfnamefont {L.~J.}\ \bibnamefont {Sham}},\ }\bibfield  {title}
  {\bibinfo {title} {{Fast Spin State Initialization in a Singly Charged
  InAs-GaAs Quantum Dot by Optical Cooling}},\ }\href
  {https://doi.org/10.1103/PhysRevLett.99.097401} {\bibfield  {journal}
  {\bibinfo  {journal} {Phys. Rev. Lett.}\ }\textbf {\bibinfo {volume} {99}},\
  \bibinfo {pages} {097401} (\bibinfo {year} {2007})}\BibitemShut {NoStop}%
\bibitem [{\citenamefont {Press}\ \emph {et~al.}(2008)\citenamefont {Press},
  \citenamefont {Ladd}, \citenamefont {Zhang},\ and\ \citenamefont
  {Yamamoto}}]{press_2008}%
  \BibitemOpen
  \bibfield  {author} {\bibinfo {author} {\bibfnamefont {D.}~\bibnamefont
  {Press}}, \bibinfo {author} {\bibfnamefont {T.~D.}\ \bibnamefont {Ladd}},
  \bibinfo {author} {\bibfnamefont {B.}~\bibnamefont {Zhang}},\ and\ \bibinfo
  {author} {\bibfnamefont {Y.}~\bibnamefont {Yamamoto}},\ }\bibfield  {title}
  {\bibinfo {title} {Complete quantum control of a single quantum dot spin
  using ultrafast optical pulses},\ }\href
  {https://doi.org/10.1038/nature07530} {\bibfield  {journal} {\bibinfo
  {journal} {Nature}\ }\textbf {\bibinfo {volume} {456}},\ \bibinfo {pages}
  {218} (\bibinfo {year} {2008})}\BibitemShut {NoStop}%
\bibitem [{\citenamefont {M{\"u}ller}\ \emph {et~al.}(2013)\citenamefont
  {M{\"u}ller}, \citenamefont {Kaldewey}, \citenamefont {Ripszam},
  \citenamefont {Wildmann}, \citenamefont {Bechtold}, \citenamefont {Bichler},
  \citenamefont {Koblm{\"u}ller}, \citenamefont {Abstreiter},\ and\
  \citenamefont {Finley}}]{muller_2013}%
  \BibitemOpen
  \bibfield  {author} {\bibinfo {author} {\bibfnamefont {K.}~\bibnamefont
  {M{\"u}ller}}, \bibinfo {author} {\bibfnamefont {T.}~\bibnamefont
  {Kaldewey}}, \bibinfo {author} {\bibfnamefont {R.}~\bibnamefont {Ripszam}},
  \bibinfo {author} {\bibfnamefont {J.~S.}\ \bibnamefont {Wildmann}}, \bibinfo
  {author} {\bibfnamefont {A.}~\bibnamefont {Bechtold}}, \bibinfo {author}
  {\bibfnamefont {M.}~\bibnamefont {Bichler}}, \bibinfo {author} {\bibfnamefont
  {G.}~\bibnamefont {Koblm{\"u}ller}}, \bibinfo {author} {\bibfnamefont
  {G.}~\bibnamefont {Abstreiter}},\ and\ \bibinfo {author} {\bibfnamefont
  {J.~J.}\ \bibnamefont {Finley}},\ }\bibfield  {title} {\bibinfo {title} {All
  optical quantum control of a spin-quantum state and ultrafast transduction
  into an electric current},\ }\href {https://doi.org/10.1038/srep01906}
  {\bibfield  {journal} {\bibinfo  {journal} {Sci. Rep.}\ }\textbf {\bibinfo
  {volume} {3}},\ \bibinfo {pages} {1906} (\bibinfo {year} {2013})}\BibitemShut
  {NoStop}%
\bibitem [{\citenamefont {Gao}\ \emph {et~al.}(2015)\citenamefont {Gao},
  \citenamefont {Imamoglu}, \citenamefont {Bernien},\ and\ \citenamefont
  {Hanson}}]{gao_2015}%
  \BibitemOpen
  \bibfield  {author} {\bibinfo {author} {\bibfnamefont {W.~B.}\ \bibnamefont
  {Gao}}, \bibinfo {author} {\bibfnamefont {A.}~\bibnamefont {Imamoglu}},
  \bibinfo {author} {\bibfnamefont {H.}~\bibnamefont {Bernien}},\ and\ \bibinfo
  {author} {\bibfnamefont {R.}~\bibnamefont {Hanson}},\ }\bibfield  {title}
  {\bibinfo {title} {Coherent manipulation, measurement and entanglement of
  individual solid-state spins using optical fields},\ }\href
  {https://doi.org/10.1038/nphoton.2015.58} {\bibfield  {journal} {\bibinfo
  {journal} {Nat. Photon.}\ }\textbf {\bibinfo {volume} {9}},\ \bibinfo {pages}
  {363} (\bibinfo {year} {2015})}\BibitemShut {NoStop}%
\bibitem [{\citenamefont {Y\ifmmode \imath~\else \i\fi{}lmaz}\ \emph
  {et~al.}(2010)\citenamefont {Y\ifmmode \imath~\else \i\fi{}lmaz},
  \citenamefont {Fallahi},\ and\ \citenamefont {Imamo\ifmmode~\breve{g}\else
  \u{g}\fi{}lu}}]{yilmaz_2010}%
  \BibitemOpen
  \bibfield  {author} {\bibinfo {author} {\bibfnamefont {S.~T.}\ \bibnamefont
  {Y\ifmmode \imath~\else \i\fi{}lmaz}}, \bibinfo {author} {\bibfnamefont
  {P.}~\bibnamefont {Fallahi}},\ and\ \bibinfo {author} {\bibfnamefont
  {A.}~\bibnamefont {Imamo\ifmmode~\breve{g}\else \u{g}\fi{}lu}},\ }\bibfield
  {title} {\bibinfo {title} {{Quantum-Dot-Spin Single-Photon Interface}},\
  }\href {https://doi.org/10.1103/PhysRevLett.105.033601} {\bibfield  {journal}
  {\bibinfo  {journal} {Phys. Rev. Lett.}\ }\textbf {\bibinfo {volume} {105}},\
  \bibinfo {pages} {033601} (\bibinfo {year} {2010})}\BibitemShut {NoStop}%
\bibitem [{\citenamefont {De~Greve}\ \emph {et~al.}(2012)\citenamefont
  {De~Greve}, \citenamefont {Yu}, \citenamefont {McMahon}, \citenamefont
  {Pelc}, \citenamefont {Natarajan}, \citenamefont {Kim}, \citenamefont {Abe},
  \citenamefont {Maier}, \citenamefont {Schneider}, \citenamefont {Kamp},
  \citenamefont {H{\"o}fling}, \citenamefont {Hadfield}, \citenamefont
  {Forchel}, \citenamefont {Fejer},\ and\ \citenamefont
  {Yamamoto}}]{degreve_2012}%
  \BibitemOpen
  \bibfield  {author} {\bibinfo {author} {\bibfnamefont {K.}~\bibnamefont
  {De~Greve}}, \bibinfo {author} {\bibfnamefont {L.}~\bibnamefont {Yu}},
  \bibinfo {author} {\bibfnamefont {P.~L.}\ \bibnamefont {McMahon}}, \bibinfo
  {author} {\bibfnamefont {J.~S.}\ \bibnamefont {Pelc}}, \bibinfo {author}
  {\bibfnamefont {C.~M.}\ \bibnamefont {Natarajan}}, \bibinfo {author}
  {\bibfnamefont {N.~Y.}\ \bibnamefont {Kim}}, \bibinfo {author} {\bibfnamefont
  {E.}~\bibnamefont {Abe}}, \bibinfo {author} {\bibfnamefont {S.}~\bibnamefont
  {Maier}}, \bibinfo {author} {\bibfnamefont {C.}~\bibnamefont {Schneider}},
  \bibinfo {author} {\bibfnamefont {M.}~\bibnamefont {Kamp}}, \bibinfo {author}
  {\bibfnamefont {S.}~\bibnamefont {H{\"o}fling}}, \bibinfo {author}
  {\bibfnamefont {R.~H.}\ \bibnamefont {Hadfield}}, \bibinfo {author}
  {\bibfnamefont {A.}~\bibnamefont {Forchel}}, \bibinfo {author} {\bibfnamefont
  {M.~M.}\ \bibnamefont {Fejer}},\ and\ \bibinfo {author} {\bibfnamefont
  {Y.}~\bibnamefont {Yamamoto}},\ }\bibfield  {title} {\bibinfo {title}
  {Quantum-dot spin--photon entanglement via frequency downconversion to
  telecom wavelength},\ }\href {https://doi.org/10.1038/nature11577} {\bibfield
   {journal} {\bibinfo  {journal} {Nature}\ }\textbf {\bibinfo {volume}
  {491}},\ \bibinfo {pages} {421} (\bibinfo {year} {2012})}\BibitemShut
  {NoStop}%
\bibitem [{\citenamefont {Debus}\ \emph
  {et~al.}(2014{\natexlab{a}})\citenamefont {Debus}, \citenamefont {Sapega},
  \citenamefont {Dunker}, \citenamefont {Yakovlev}, \citenamefont {Reuter},
  \citenamefont {Wieck},\ and\ \citenamefont {Bayer}}]{debus_2014_2}%
  \BibitemOpen
  \bibfield  {author} {\bibinfo {author} {\bibfnamefont {J.}~\bibnamefont
  {Debus}}, \bibinfo {author} {\bibfnamefont {V.~F.}\ \bibnamefont {Sapega}},
  \bibinfo {author} {\bibfnamefont {D.}~\bibnamefont {Dunker}}, \bibinfo
  {author} {\bibfnamefont {D.~R.}\ \bibnamefont {Yakovlev}}, \bibinfo {author}
  {\bibfnamefont {D.}~\bibnamefont {Reuter}}, \bibinfo {author} {\bibfnamefont
  {A.~D.}\ \bibnamefont {Wieck}},\ and\ \bibinfo {author} {\bibfnamefont
  {M.}~\bibnamefont {Bayer}},\ }\bibfield  {title} {\bibinfo {title}
  {{Spin-flip Raman scattering of the resident electron in singly charged
  (In,Ga)As/GaAs quantum dot ensembles}},\ }\href
  {https://doi.org/10.1103/PhysRevB.90.235404} {\bibfield  {journal} {\bibinfo
  {journal} {Phys. Rev. B}\ }\textbf {\bibinfo {volume} {90}},\ \bibinfo
  {pages} {235404} (\bibinfo {year} {2014}{\natexlab{a}})}\BibitemShut
  {NoStop}%
\bibitem [{\citenamefont {Dreiser}\ \emph {et~al.}(2008)\citenamefont
  {Dreiser}, \citenamefont {Atat\"ure}, \citenamefont {Galland}, \citenamefont
  {M\"uller}, \citenamefont {Badolato},\ and\ \citenamefont
  {Imamoglu}}]{dreiser_2008}%
  \BibitemOpen
  \bibfield  {author} {\bibinfo {author} {\bibfnamefont {J.}~\bibnamefont
  {Dreiser}}, \bibinfo {author} {\bibfnamefont {M.}~\bibnamefont {Atat\"ure}},
  \bibinfo {author} {\bibfnamefont {C.}~\bibnamefont {Galland}}, \bibinfo
  {author} {\bibfnamefont {T.}~\bibnamefont {M\"uller}}, \bibinfo {author}
  {\bibfnamefont {A.}~\bibnamefont {Badolato}},\ and\ \bibinfo {author}
  {\bibfnamefont {A.}~\bibnamefont {Imamoglu}},\ }\bibfield  {title} {\bibinfo
  {title} {Optical investigations of quantum dot spin dynamics as a function of
  external electric and magnetic fields},\ }\href
  {https://doi.org/10.1103/PhysRevB.77.075317} {\bibfield  {journal} {\bibinfo
  {journal} {Phys. Rev. B}\ }\textbf {\bibinfo {volume} {77}},\ \bibinfo
  {pages} {075317} (\bibinfo {year} {2008})}\BibitemShut {NoStop}%
\bibitem [{\citenamefont {Atat\"ure}\ \emph {et~al.}(2006)\citenamefont
  {Atat\"ure}, \citenamefont {Dreiser}, \citenamefont {Badolato}, \citenamefont
  {H\"ogele}, \citenamefont {Karrai},\ and\ \citenamefont
  {Imamoglu}}]{atatuere_2006}%
  \BibitemOpen
  \bibfield  {author} {\bibinfo {author} {\bibfnamefont {M.}~\bibnamefont
  {Atat\"ure}}, \bibinfo {author} {\bibfnamefont {J.}~\bibnamefont {Dreiser}},
  \bibinfo {author} {\bibfnamefont {A.}~\bibnamefont {Badolato}}, \bibinfo
  {author} {\bibfnamefont {A.}~\bibnamefont {H\"ogele}}, \bibinfo {author}
  {\bibfnamefont {K.}~\bibnamefont {Karrai}},\ and\ \bibinfo {author}
  {\bibfnamefont {A.}~\bibnamefont {Imamoglu}},\ }\bibfield  {title} {\bibinfo
  {title} {Quantum-dot spin-state preparation with near-unity fidelity},\
  }\href {https://doi.org/10.1126/science.1126074} {\bibfield  {journal}
  {\bibinfo  {journal} {Science}\ }\textbf {\bibinfo {volume} {312}},\ \bibinfo
  {pages} {551} (\bibinfo {year} {2006})}\BibitemShut {NoStop}%
\bibitem [{\citenamefont {Mannel}\ \emph {et~al.}()\citenamefont {Mannel},
  \citenamefont {Kerski}, \citenamefont {Lochner}, \citenamefont {Z{\"o}llner},
  \citenamefont {Wieck}, \citenamefont {Ludwig}, \citenamefont {Lorke},\ and\
  \citenamefont {Geller}}]{mannel_2021}%
  \BibitemOpen
  \bibfield  {author} {\bibinfo {author} {\bibfnamefont {H.}~\bibnamefont
  {Mannel}}, \bibinfo {author} {\bibfnamefont {J.}~\bibnamefont {Kerski}},
  \bibinfo {author} {\bibfnamefont {P.}~\bibnamefont {Lochner}}, \bibinfo
  {author} {\bibfnamefont {M.}~\bibnamefont {Z{\"o}llner}}, \bibinfo {author}
  {\bibfnamefont {A.~D.}\ \bibnamefont {Wieck}}, \bibinfo {author}
  {\bibfnamefont {A.}~\bibnamefont {Ludwig}}, \bibinfo {author} {\bibfnamefont
  {A.}~\bibnamefont {Lorke}},\ and\ \bibinfo {author} {\bibfnamefont
  {M.}~\bibnamefont {Geller}},\ }\bibfield  {title} {\bibinfo {title} {Auger
  and spin dynamics in a self-assembled quantum dot},\ }\href
  {https://arxiv.org/abs/2110.12213} {\bibinfo  {journal} {arXiv:2110.12213}\
  }\BibitemShut {NoStop}%
\bibitem [{\citenamefont {Kurzmann}\ \emph {et~al.}(2019)\citenamefont
  {Kurzmann}, \citenamefont {Stegmann}, \citenamefont {Kerski}, \citenamefont
  {Schott}, \citenamefont {Ludwig}, \citenamefont {Wieck}, \citenamefont
  {K\"onig}, \citenamefont {Lorke},\ and\ \citenamefont
  {Geller}}]{kurzmann2019optical}%
  \BibitemOpen
\bibfield  {journal} {  }\bibfield  {author} {\bibinfo {author} {\bibfnamefont
  {A.}~\bibnamefont {Kurzmann}}, \bibinfo {author} {\bibfnamefont
  {P.}~\bibnamefont {Stegmann}}, \bibinfo {author} {\bibfnamefont
  {J.}~\bibnamefont {Kerski}}, \bibinfo {author} {\bibfnamefont
  {R.}~\bibnamefont {Schott}}, \bibinfo {author} {\bibfnamefont
  {A.}~\bibnamefont {Ludwig}}, \bibinfo {author} {\bibfnamefont {A.~D.}\
  \bibnamefont {Wieck}}, \bibinfo {author} {\bibfnamefont {J.}~\bibnamefont
  {K\"onig}}, \bibinfo {author} {\bibfnamefont {A.}~\bibnamefont {Lorke}},\
  and\ \bibinfo {author} {\bibfnamefont {M.}~\bibnamefont {Geller}},\
  }\bibfield  {title} {\bibinfo {title} {Optical {D}etection of
  {S}ingle-{E}lectron {T}unneling into a {S}emiconductor {Q}uantum {D}ot},\
  }\href {https://doi.org/10.1103/PhysRevLett.122.247403} {\bibfield  {journal}
  {\bibinfo  {journal} {Phys. Rev. Lett.}\ }\textbf {\bibinfo {volume} {122}},\
  \bibinfo {pages} {247403} (\bibinfo {year} {2019})}\BibitemShut {NoStop}%
\bibitem [{\citenamefont {Lochner}\ \emph {et~al.}(2020)\citenamefont
  {Lochner}, \citenamefont {Kurzmann}, \citenamefont {Kerski}, \citenamefont
  {Stegmann}, \citenamefont {K{\"o}nig}, \citenamefont {Wieck}, \citenamefont
  {Ludwig}, \citenamefont {Lorke},\ and\ \citenamefont
  {Geller}}]{lochner_2020}%
  \BibitemOpen
  \bibfield  {author} {\bibinfo {author} {\bibfnamefont {P.}~\bibnamefont
  {Lochner}}, \bibinfo {author} {\bibfnamefont {A.}~\bibnamefont {Kurzmann}},
  \bibinfo {author} {\bibfnamefont {J.}~\bibnamefont {Kerski}}, \bibinfo
  {author} {\bibfnamefont {P.}~\bibnamefont {Stegmann}}, \bibinfo {author}
  {\bibfnamefont {J.}~\bibnamefont {K{\"o}nig}}, \bibinfo {author}
  {\bibfnamefont {A.~D.}\ \bibnamefont {Wieck}}, \bibinfo {author}
  {\bibfnamefont {A.}~\bibnamefont {Ludwig}}, \bibinfo {author} {\bibfnamefont
  {A.}~\bibnamefont {Lorke}},\ and\ \bibinfo {author} {\bibfnamefont
  {M.}~\bibnamefont {Geller}},\ }\bibfield  {title} {\bibinfo {title}
  {Real-{T}ime {D}etection of {S}ingle {A}uger {R}ecombination {E}vents in a
  {S}elf-{A}ssembled {Q}uantum {D}ot},\ }\href
  {https://doi.org/10.1021/acs.nanolett.9b04650} {\bibfield  {journal}
  {\bibinfo  {journal} {Nano Lett.}\ }\textbf {\bibinfo {volume} {20}},\
  \bibinfo {pages} {1631} (\bibinfo {year} {2020})}\BibitemShut {NoStop}%
\bibitem [{\citenamefont {Gillard}\ \emph {et~al.}(2021)\citenamefont
  {Gillard}, \citenamefont {Griffiths}, \citenamefont {Ragunathan},
  \citenamefont {Ulhaq}, \citenamefont {McEwan}, \citenamefont {Clarke},\ and\
  \citenamefont {Chekhovich}}]{gillard_2021}%
  \BibitemOpen
  \bibfield  {author} {\bibinfo {author} {\bibfnamefont {G.}~\bibnamefont
  {Gillard}}, \bibinfo {author} {\bibfnamefont {I.~M.}\ \bibnamefont
  {Griffiths}}, \bibinfo {author} {\bibfnamefont {G.}~\bibnamefont
  {Ragunathan}}, \bibinfo {author} {\bibfnamefont {A.}~\bibnamefont {Ulhaq}},
  \bibinfo {author} {\bibfnamefont {C.}~\bibnamefont {McEwan}}, \bibinfo
  {author} {\bibfnamefont {E.}~\bibnamefont {Clarke}},\ and\ \bibinfo {author}
  {\bibfnamefont {E.~A.}\ \bibnamefont {Chekhovich}},\ }\bibfield  {title}
  {\bibinfo {title} {Fundamental limits of electron and nuclear spin qubit
  lifetimes in an isolated self-assembled quantum dot},\ }\href
  {https://doi.org/10.1038/s41534-021-00378-2} {\bibfield  {journal} {\bibinfo
  {journal} {Npj Quantum Inf.}\ }\textbf {\bibinfo {volume} {7}},\ \bibinfo
  {pages} {43} (\bibinfo {year} {2021})}\BibitemShut {NoStop}%
\bibitem [{\citenamefont {Dahbashi}\ \emph {et~al.}(2014)\citenamefont
  {Dahbashi}, \citenamefont {H\"ubner}, \citenamefont {Berski}, \citenamefont
  {Pierz},\ and\ \citenamefont {Oestreich}}]{dahbashi_2014}%
  \BibitemOpen
  \bibfield  {author} {\bibinfo {author} {\bibfnamefont {R.}~\bibnamefont
  {Dahbashi}}, \bibinfo {author} {\bibfnamefont {J.}~\bibnamefont {H\"ubner}},
  \bibinfo {author} {\bibfnamefont {F.}~\bibnamefont {Berski}}, \bibinfo
  {author} {\bibfnamefont {K.}~\bibnamefont {Pierz}},\ and\ \bibinfo {author}
  {\bibfnamefont {M.}~\bibnamefont {Oestreich}},\ }\bibfield  {title} {\bibinfo
  {title} {{Optical Spin Noise of a Single Hole Spin Localized in an (InGa)As
  Quantum Dot}},\ }\href {https://doi.org/10.1103/PhysRevLett.112.156601}
  {\bibfield  {journal} {\bibinfo  {journal} {Phys. Rev. Lett.}\ }\textbf
  {\bibinfo {volume} {112}},\ \bibinfo {pages} {156601} (\bibinfo {year}
  {2014})}\BibitemShut {NoStop}%
\bibitem [{\citenamefont {Merkulov}\ \emph {et~al.}(2002)\citenamefont
  {Merkulov}, \citenamefont {Efros},\ and\ \citenamefont
  {Rosen}}]{merkulov_2002}%
  \BibitemOpen
  \bibfield  {author} {\bibinfo {author} {\bibfnamefont {I.~A.}\ \bibnamefont
  {Merkulov}}, \bibinfo {author} {\bibfnamefont {A.~L.}\ \bibnamefont
  {Efros}},\ and\ \bibinfo {author} {\bibfnamefont {M.}~\bibnamefont {Rosen}},\
  }\bibfield  {title} {\bibinfo {title} {Electron spin relaxation by nuclei in
  semiconductor quantum dots},\ }\href
  {https://doi.org/10.1103/PhysRevB.65.205309} {\bibfield  {journal} {\bibinfo
  {journal} {Phys. Rev. B}\ }\textbf {\bibinfo {volume} {65}},\ \bibinfo
  {pages} {205309} (\bibinfo {year} {2002})}\BibitemShut {NoStop}%
\bibitem [{\citenamefont {Bracker}\ \emph {et~al.}(2005)\citenamefont
  {Bracker}, \citenamefont {Stinaff}, \citenamefont {Gammon}, \citenamefont
  {Ware}, \citenamefont {Tischler}, \citenamefont {Shabaev}, \citenamefont
  {Efros}, \citenamefont {Park}, \citenamefont {Gershoni}, \citenamefont
  {Korenev},\ and\ \citenamefont {Merkulov}}]{bracker_2005}%
  \BibitemOpen
  \bibfield  {author} {\bibinfo {author} {\bibfnamefont {A.~S.}\ \bibnamefont
  {Bracker}}, \bibinfo {author} {\bibfnamefont {E.~A.}\ \bibnamefont
  {Stinaff}}, \bibinfo {author} {\bibfnamefont {D.}~\bibnamefont {Gammon}},
  \bibinfo {author} {\bibfnamefont {M.~E.}\ \bibnamefont {Ware}}, \bibinfo
  {author} {\bibfnamefont {J.~G.}\ \bibnamefont {Tischler}}, \bibinfo {author}
  {\bibfnamefont {A.}~\bibnamefont {Shabaev}}, \bibinfo {author} {\bibfnamefont
  {A.~L.}\ \bibnamefont {Efros}}, \bibinfo {author} {\bibfnamefont
  {D.}~\bibnamefont {Park}}, \bibinfo {author} {\bibfnamefont {D.}~\bibnamefont
  {Gershoni}}, \bibinfo {author} {\bibfnamefont {V.~L.}\ \bibnamefont
  {Korenev}},\ and\ \bibinfo {author} {\bibfnamefont {I.~A.}\ \bibnamefont
  {Merkulov}},\ }\bibfield  {title} {\bibinfo {title} {{Optical Pumping of the
  Electronic and Nuclear Spin of Single Charge-Tunable Quantum Dots}},\ }\href
  {https://doi.org/10.1103/PhysRevLett.94.047402} {\bibfield  {journal}
  {\bibinfo  {journal} {Phys. Rev. Lett.}\ }\textbf {\bibinfo {volume} {94}},\
  \bibinfo {pages} {047402} (\bibinfo {year} {2005})}\BibitemShut {NoStop}%
\bibitem [{\citenamefont {Kuhlmann}\ \emph {et~al.}(2013)\citenamefont
  {Kuhlmann}, \citenamefont {Houel}, \citenamefont {Ludwig}, \citenamefont
  {Greuter}, \citenamefont {Reuter}, \citenamefont {Wieck}, \citenamefont
  {Poggio},\ and\ \citenamefont {Warburton}}]{kuhlmann_2013}%
  \BibitemOpen
  \bibfield  {author} {\bibinfo {author} {\bibfnamefont {A.~V.}\ \bibnamefont
  {Kuhlmann}}, \bibinfo {author} {\bibfnamefont {J.}~\bibnamefont {Houel}},
  \bibinfo {author} {\bibfnamefont {A.}~\bibnamefont {Ludwig}}, \bibinfo
  {author} {\bibfnamefont {L.}~\bibnamefont {Greuter}}, \bibinfo {author}
  {\bibfnamefont {D.}~\bibnamefont {Reuter}}, \bibinfo {author} {\bibfnamefont
  {A.~D.}\ \bibnamefont {Wieck}}, \bibinfo {author} {\bibfnamefont
  {M.}~\bibnamefont {Poggio}},\ and\ \bibinfo {author} {\bibfnamefont {R.~J.}\
  \bibnamefont {Warburton}},\ }\bibfield  {title} {\bibinfo {title} {Charge
  noise and spin noise in a semiconductor quantum device},\ }\href
  {https://doi.org/10.1038/nphys2688} {\bibfield  {journal} {\bibinfo
  {journal} {Nat. Phys.}\ }\textbf {\bibinfo {volume} {9}},\ \bibinfo {pages}
  {570} (\bibinfo {year} {2013})}\BibitemShut {NoStop}%
\bibitem [{\citenamefont {Khaetskii}\ and\ \citenamefont
  {Nazarov}(2001)}]{khaetskii_2001}%
  \BibitemOpen
  \bibfield  {author} {\bibinfo {author} {\bibfnamefont {A.~V.}\ \bibnamefont
  {Khaetskii}}\ and\ \bibinfo {author} {\bibfnamefont {Y.~V.}\ \bibnamefont
  {Nazarov}},\ }\bibfield  {title} {\bibinfo {title} {Spin-flip transitions
  between {Z}eeman sublevels in semiconductor quantum dots},\ }\href
  {https://doi.org/10.1103/PhysRevB.64.125316} {\bibfield  {journal} {\bibinfo
  {journal} {Phys. Rev. B}\ }\textbf {\bibinfo {volume} {64}},\ \bibinfo
  {pages} {125316} (\bibinfo {year} {2001})}\BibitemShut {NoStop}%
\bibitem [{\citenamefont {Kurzmann}\ \emph
  {et~al.}(2016{\natexlab{a}})\citenamefont {Kurzmann}, \citenamefont {Ludwig},
  \citenamefont {Wieck}, \citenamefont {Lorke},\ and\ \citenamefont
  {Geller}}]{kurzmann_2016_auger}%
  \BibitemOpen
  \bibfield  {author} {\bibinfo {author} {\bibfnamefont {A.}~\bibnamefont
  {Kurzmann}}, \bibinfo {author} {\bibfnamefont {A.}~\bibnamefont {Ludwig}},
  \bibinfo {author} {\bibfnamefont {A.~D.}\ \bibnamefont {Wieck}}, \bibinfo
  {author} {\bibfnamefont {A.}~\bibnamefont {Lorke}},\ and\ \bibinfo {author}
  {\bibfnamefont {M.}~\bibnamefont {Geller}},\ }\bibfield  {title} {\bibinfo
  {title} {Auger {R}ecombination in {S}elf-{A}ssembled {Q}uantum {D}ots:
  {Q}uenching and {B}roadening of the {C}harged {E}xciton {T}ransition},\
  }\href {https://doi.org/10.1021/acs.nanolett.6b01082} {\bibfield  {journal}
  {\bibinfo  {journal} {Nano Lett.}\ }\textbf {\bibinfo {volume} {16}},\
  \bibinfo {pages} {3367} (\bibinfo {year} {2016}{\natexlab{a}})}\BibitemShut
  {NoStop}%
\bibitem [{\citenamefont {L{\"o}bl}\ \emph {et~al.}(2020)\citenamefont
  {L{\"o}bl}, \citenamefont {Spinnler}, \citenamefont {Javadi}, \citenamefont
  {Zhai}, \citenamefont {Nguyen}, \citenamefont {Ritzmann}, \citenamefont
  {Midolo}, \citenamefont {Lodahl}, \citenamefont {Wieck}, \citenamefont
  {Ludwig},\ and\ \citenamefont {Warburton}}]{lobl_2020}%
  \BibitemOpen
  \bibfield  {author} {\bibinfo {author} {\bibfnamefont {M.~C.}\ \bibnamefont
  {L{\"o}bl}}, \bibinfo {author} {\bibfnamefont {C.}~\bibnamefont {Spinnler}},
  \bibinfo {author} {\bibfnamefont {A.}~\bibnamefont {Javadi}}, \bibinfo
  {author} {\bibfnamefont {L.}~\bibnamefont {Zhai}}, \bibinfo {author}
  {\bibfnamefont {G.~N.}\ \bibnamefont {Nguyen}}, \bibinfo {author}
  {\bibfnamefont {J.}~\bibnamefont {Ritzmann}}, \bibinfo {author}
  {\bibfnamefont {L.}~\bibnamefont {Midolo}}, \bibinfo {author} {\bibfnamefont
  {P.}~\bibnamefont {Lodahl}}, \bibinfo {author} {\bibfnamefont {A.~D.}\
  \bibnamefont {Wieck}}, \bibinfo {author} {\bibfnamefont {A.}~\bibnamefont
  {Ludwig}},\ and\ \bibinfo {author} {\bibfnamefont {R.~J.}\ \bibnamefont
  {Warburton}},\ }\bibfield  {title} {\bibinfo {title} {Radiative {A}uger
  process in the single-photon limit},\ }\href
  {https://doi.org/10.1038/s41565-020-0697-2} {\bibfield  {journal} {\bibinfo
  {journal} {Nat. Nanotechnol.}\ }\textbf {\bibinfo {volume} {15}},\ \bibinfo
  {pages} {558} (\bibinfo {year} {2020})}\BibitemShut {NoStop}%
\bibitem [{\citenamefont {Brandes}(2008)}]{brandes_2008}%
  \BibitemOpen
  \bibfield  {author} {\bibinfo {author} {\bibfnamefont {T.}~\bibnamefont
  {Brandes}},\ }\bibfield  {title} {\bibinfo {title} {Waiting times and noise
  in single particle transport},\ }\href
  {https://doi.org/https://doi.org/10.1002/andp.20085200707} {\bibfield
  {journal} {\bibinfo  {journal} {Ann. Phys.}\ }\textbf {\bibinfo {volume}
  {520}},\ \bibinfo {pages} {477} (\bibinfo {year} {2008})}\BibitemShut
  {NoStop}%
\bibitem [{\citenamefont {Albert}\ \emph {et~al.}(2012)\citenamefont {Albert},
  \citenamefont {Haack}, \citenamefont {Flindt},\ and\ \citenamefont
  {B\"uttiker}}]{albert_2012}%
  \BibitemOpen
  \bibfield  {author} {\bibinfo {author} {\bibfnamefont {M.}~\bibnamefont
  {Albert}}, \bibinfo {author} {\bibfnamefont {G.}~\bibnamefont {Haack}},
  \bibinfo {author} {\bibfnamefont {C.}~\bibnamefont {Flindt}},\ and\ \bibinfo
  {author} {\bibfnamefont {M.}~\bibnamefont {B\"uttiker}},\ }\bibfield  {title}
  {\bibinfo {title} {Electron {W}aiting {T}imes in {M}esoscopic {C}onductors},\
  }\href {https://doi.org/10.1103/PhysRevLett.108.186806} {\bibfield  {journal}
  {\bibinfo  {journal} {Phys. Rev. Lett.}\ }\textbf {\bibinfo {volume} {108}},\
  \bibinfo {pages} {186806} (\bibinfo {year} {2012})}\BibitemShut {NoStop}%
\bibitem [{\citenamefont {Rudge}\ and\ \citenamefont
  {Kosov}(2018)}]{rudge_2018}%
  \BibitemOpen
  \bibfield  {author} {\bibinfo {author} {\bibfnamefont {S.~L.}\ \bibnamefont
  {Rudge}}\ and\ \bibinfo {author} {\bibfnamefont {D.~S.}\ \bibnamefont
  {Kosov}},\ }\bibfield  {title} {\bibinfo {title} {Distribution of waiting
  times between electron cotunneling events},\ }\href
  {https://doi.org/10.1103/PhysRevB.98.245402} {\bibfield  {journal} {\bibinfo
  {journal} {Phys. Rev. B}\ }\textbf {\bibinfo {volume} {98}},\ \bibinfo
  {pages} {245402} (\bibinfo {year} {2018})}\BibitemShut {NoStop}%
\bibitem [{\citenamefont {Kleinherbers}\ \emph {et~al.}(2021)\citenamefont
  {Kleinherbers}, \citenamefont {Stegmann},\ and\ \citenamefont
  {K\"onig}}]{kleinherbers_2021_synchro}%
  \BibitemOpen
  \bibfield  {author} {\bibinfo {author} {\bibfnamefont {E.}~\bibnamefont
  {Kleinherbers}}, \bibinfo {author} {\bibfnamefont {P.}~\bibnamefont
  {Stegmann}},\ and\ \bibinfo {author} {\bibfnamefont {J.}~\bibnamefont
  {K\"onig}},\ }\bibfield  {title} {\bibinfo {title} {Synchronized coherent
  charge oscillations in coupled double quantum dots},\ }\href
  {https://doi.org/10.1103/PhysRevB.104.165304} {\bibfield  {journal} {\bibinfo
   {journal} {Phys. Rev. B}\ }\textbf {\bibinfo {volume} {104}},\ \bibinfo
  {pages} {165304} (\bibinfo {year} {2021})}\BibitemShut {NoStop}%
\bibitem [{\citenamefont {Levitov}\ and\ \citenamefont
  {Lesovik}(1993)}]{levitov_1993}%
  \BibitemOpen
  \bibfield  {author} {\bibinfo {author} {\bibfnamefont {L.~S.}\ \bibnamefont
  {Levitov}}\ and\ \bibinfo {author} {\bibfnamefont {G.~B.}\ \bibnamefont
  {Lesovik}},\ }\bibfield  {title} {\bibinfo {title} {Charge distribution in
  quantum shot noise},\ }\href@noop {} {\bibfield  {journal} {\bibinfo
  {journal} {JETP Lett.}\ }\textbf {\bibinfo {volume} {58}},\ \bibinfo {pages}
  {230} (\bibinfo {year} {1993})}\BibitemShut {NoStop}%
\bibitem [{\citenamefont {Blanter}\ and\ \citenamefont
  {B{\"u}ttiker}(2000)}]{blanter_2000}%
  \BibitemOpen
  \bibfield  {author} {\bibinfo {author} {\bibfnamefont {Y.}~\bibnamefont
  {Blanter}}\ and\ \bibinfo {author} {\bibfnamefont {M.}~\bibnamefont
  {B{\"u}ttiker}},\ }\bibfield  {title} {\bibinfo {title} {Shot noise in
  mesoscopic conductors},\ }\href
  {https://doi.org/10.1016/S0370-1573(99)00123-4} {\bibfield  {journal}
  {\bibinfo  {journal} {Phys. Rep.}\ }\textbf {\bibinfo {volume} {336}},\
  \bibinfo {pages} {1} (\bibinfo {year} {2000})}\BibitemShut {NoStop}%
\bibitem [{\citenamefont {Marcos}\ \emph {et~al.}(2010)\citenamefont {Marcos},
  \citenamefont {Emary}, \citenamefont {Brandes},\ and\ \citenamefont
  {Aguado}}]{marcos_2010}%
  \BibitemOpen
  \bibfield  {author} {\bibinfo {author} {\bibfnamefont {D.}~\bibnamefont
  {Marcos}}, \bibinfo {author} {\bibfnamefont {C.}~\bibnamefont {Emary}},
  \bibinfo {author} {\bibfnamefont {T.}~\bibnamefont {Brandes}},\ and\ \bibinfo
  {author} {\bibfnamefont {R.}~\bibnamefont {Aguado}},\ }\bibfield  {title}
  {\bibinfo {title} {{Finite-frequency counting statistics of electron
  transport: Markovian theory}},\ }\href
  {https://doi.org/10.1088/1367-2630/12/12/123009} {\bibfield  {journal}
  {\bibinfo  {journal} {New J. Phys.}\ }\textbf {\bibinfo {volume} {12}},\
  \bibinfo {pages} {123009} (\bibinfo {year} {2010})}\BibitemShut {NoStop}%
\bibitem [{\citenamefont {Rudge}\ and\ \citenamefont
  {Kosov}(2019)}]{rudge_2019}%
  \BibitemOpen
  \bibfield  {author} {\bibinfo {author} {\bibfnamefont {S.~L.}\ \bibnamefont
  {Rudge}}\ and\ \bibinfo {author} {\bibfnamefont {D.~S.}\ \bibnamefont
  {Kosov}},\ }\bibfield  {title} {\bibinfo {title} {Counting quantum jumps: {A}
  summary and comparison of fixed-time and fluctuating-time statistics in
  electron transport},\ }\href {https://doi.org/10.1063/1.5108518} {\bibfield
  {journal} {\bibinfo  {journal} {J. Chem. Phys.}\ }\textbf {\bibinfo {volume}
  {151}},\ \bibinfo {pages} {034107} (\bibinfo {year} {2019})}\BibitemShut
  {NoStop}%
\bibitem [{\citenamefont {Beenakker}\ and\ \citenamefont
  {Schomerus}(2001)}]{beenakker_counting_2001}%
  \BibitemOpen
  \bibfield  {author} {\bibinfo {author} {\bibfnamefont {C.~W.~J.}\
  \bibnamefont {Beenakker}}\ and\ \bibinfo {author} {\bibfnamefont
  {H.}~\bibnamefont {Schomerus}},\ }\bibfield  {title} {\bibinfo {title}
  {Counting {S}tatistics of {P}hotons {P}roduced by {E}lectronic {S}hot
  {N}oise},\ }\href {https://doi.org/10.1103/PhysRevLett.86.700} {\bibfield
  {journal} {\bibinfo  {journal} {Phys. Rev. Lett.}\ }\textbf {\bibinfo
  {volume} {86}},\ \bibinfo {pages} {700} (\bibinfo {year} {2001})}\BibitemShut
  {NoStop}%
\bibitem [{\citenamefont {Flindt}\ \emph {et~al.}(2009)\citenamefont {Flindt},
  \citenamefont {Fricke}, \citenamefont {Hohls}, \citenamefont {Novotn{\'y}},
  \citenamefont {Neto{\v c}n{\'y}}, \citenamefont {Brandes},\ and\
  \citenamefont {Haug}}]{flindt2009universal}%
  \BibitemOpen
  \bibfield  {author} {\bibinfo {author} {\bibfnamefont {C.}~\bibnamefont
  {Flindt}}, \bibinfo {author} {\bibfnamefont {C.}~\bibnamefont {Fricke}},
  \bibinfo {author} {\bibfnamefont {F.}~\bibnamefont {Hohls}}, \bibinfo
  {author} {\bibfnamefont {T.}~\bibnamefont {Novotn{\'y}}}, \bibinfo {author}
  {\bibfnamefont {K.}~\bibnamefont {Neto{\v c}n{\'y}}}, \bibinfo {author}
  {\bibfnamefont {T.}~\bibnamefont {Brandes}},\ and\ \bibinfo {author}
  {\bibfnamefont {R.~J.}\ \bibnamefont {Haug}},\ }\bibfield  {title} {\bibinfo
  {title} {Universal oscillations in counting statistics},\ }\href
  {https://doi.org/10.1073/pnas.0901002106} {\bibfield  {journal} {\bibinfo
  {journal} {Pro. Natl. Acad. Sci. USA}\ }\textbf {\bibinfo {volume} {106}},\
  \bibinfo {pages} {10116} (\bibinfo {year} {2009})}\BibitemShut {NoStop}%
\bibitem [{\citenamefont {Kambly}\ \emph {et~al.}(2011)\citenamefont {Kambly},
  \citenamefont {Flindt},\ and\ \citenamefont {B\"uttiker}}]{kambly_2011}%
  \BibitemOpen
  \bibfield  {author} {\bibinfo {author} {\bibfnamefont {D.}~\bibnamefont
  {Kambly}}, \bibinfo {author} {\bibfnamefont {C.}~\bibnamefont {Flindt}},\
  and\ \bibinfo {author} {\bibfnamefont {M.}~\bibnamefont {B\"uttiker}},\
  }\bibfield  {title} {\bibinfo {title} {Factorial cumulants reveal
  interactions in counting statistics},\ }\href
  {https://doi.org/10.1103/PhysRevB.83.075432} {\bibfield  {journal} {\bibinfo
  {journal} {Phys. Rev. B}\ }\textbf {\bibinfo {volume} {83}},\ \bibinfo
  {pages} {075432} (\bibinfo {year} {2011})}\BibitemShut {NoStop}%
\bibitem [{\citenamefont {Stegmann}\ \emph {et~al.}(2015)\citenamefont
  {Stegmann}, \citenamefont {Sothmann}, \citenamefont {Hucht},\ and\
  \citenamefont {K\"onig}}]{stegmann2015detection}%
  \BibitemOpen
  \bibfield  {author} {\bibinfo {author} {\bibfnamefont {P.}~\bibnamefont
  {Stegmann}}, \bibinfo {author} {\bibfnamefont {B.}~\bibnamefont {Sothmann}},
  \bibinfo {author} {\bibfnamefont {A.}~\bibnamefont {Hucht}},\ and\ \bibinfo
  {author} {\bibfnamefont {J.}~\bibnamefont {K\"onig}},\ }\bibfield  {title}
  {\bibinfo {title} {Detection of interactions via generalized factorial
  cumulants in systems in and out of equilibrium},\ }\href
  {https://doi.org/10.1103/PhysRevB.92.155413} {\bibfield  {journal} {\bibinfo
  {journal} {Phys. Rev. B}\ }\textbf {\bibinfo {volume} {92}},\ \bibinfo
  {pages} {155413} (\bibinfo {year} {2015})}\BibitemShut {NoStop}%
\bibitem [{\citenamefont {Stegmann}\ and\ \citenamefont
  {K\"onig}(2016)}]{stegmann2016short}%
  \BibitemOpen
  \bibfield  {author} {\bibinfo {author} {\bibfnamefont {P.}~\bibnamefont
  {Stegmann}}\ and\ \bibinfo {author} {\bibfnamefont {J.}~\bibnamefont
  {K\"onig}},\ }\bibfield  {title} {\bibinfo {title} {Short-time counting
  statistics of charge transfer in {C}oulomb-blockade systems},\ }\href
  {https://doi.org/10.1103/PhysRevB.94.125433} {\bibfield  {journal} {\bibinfo
  {journal} {Phys. Rev. B}\ }\textbf {\bibinfo {volume} {94}},\ \bibinfo
  {pages} {125433} (\bibinfo {year} {2016})}\BibitemShut {NoStop}%
\bibitem [{\citenamefont {Kleinherbers}\ \emph {et~al.}(2018)\citenamefont
  {Kleinherbers}, \citenamefont {Stegmann},\ and\ \citenamefont
  {K{\"o}nig}}]{kleinherbers2018revealing}%
  \BibitemOpen
  \bibfield  {author} {\bibinfo {author} {\bibfnamefont {E.}~\bibnamefont
  {Kleinherbers}}, \bibinfo {author} {\bibfnamefont {P.}~\bibnamefont
  {Stegmann}},\ and\ \bibinfo {author} {\bibfnamefont {J.}~\bibnamefont
  {K{\"o}nig}},\ }\bibfield  {title} {\bibinfo {title} {Revealing attractive
  electron--electron interaction in a quantum dot by full counting
  statistics},\ }\href {https://doi.org/10.1088/1367-2630/aad14a} {\bibfield
  {journal} {\bibinfo  {journal} {New J. Phys.}\ }\textbf {\bibinfo {volume}
  {20}},\ \bibinfo {pages} {073023} (\bibinfo {year} {2018})}\BibitemShut
  {NoStop}%
\bibitem [{\citenamefont {K{\"o}nig}\ and\ \citenamefont
  {Hucht}(2021)}]{koenig_2021}%
  \BibitemOpen
  \bibfield  {author} {\bibinfo {author} {\bibfnamefont {J.}~\bibnamefont
  {K{\"o}nig}}\ and\ \bibinfo {author} {\bibfnamefont {A.}~\bibnamefont
  {Hucht}},\ }\bibfield  {title} {\bibinfo {title} {{Newton series expansion of
  bosonic operator functions}},\ }\href
  {https://doi.org/10.21468/SciPostPhys.10.1.007} {\bibfield  {journal}
  {\bibinfo  {journal} {SciPost Phys.}\ }\textbf {\bibinfo {volume} {10}},\
  \bibinfo {pages} {7} (\bibinfo {year} {2021})}\BibitemShut {NoStop}%
\bibitem [{\citenamefont {Kleinherbers}\ \emph {et~al.}(2022)\citenamefont
  {Kleinherbers}, \citenamefont {Stegmann}, \citenamefont {Kurzmann},
  \citenamefont {Geller}, \citenamefont {Lorke},\ and\ \citenamefont
  {K\"onig}}]{kleinherbers_2021_pushing}%
  \BibitemOpen
  \bibfield  {author} {\bibinfo {author} {\bibfnamefont {E.}~\bibnamefont
  {Kleinherbers}}, \bibinfo {author} {\bibfnamefont {P.}~\bibnamefont
  {Stegmann}}, \bibinfo {author} {\bibfnamefont {A.}~\bibnamefont {Kurzmann}},
  \bibinfo {author} {\bibfnamefont {M.}~\bibnamefont {Geller}}, \bibinfo
  {author} {\bibfnamefont {A.}~\bibnamefont {Lorke}},\ and\ \bibinfo {author}
  {\bibfnamefont {J.}~\bibnamefont {K\"onig}},\ }\bibfield  {title} {\bibinfo
  {title} {Pushing the {L}imits in {R}eal-{T}ime {M}easurements of {Q}uantum
  {D}ynamics},\ }\href {https://doi.org/10.1103/PhysRevLett.128.087701}
  {\bibfield  {journal} {\bibinfo  {journal} {Phys. Rev. Lett.}\ }\textbf
  {\bibinfo {volume} {128}},\ \bibinfo {pages} {087701} (\bibinfo {year}
  {2022})}\BibitemShut {NoStop}%
\bibitem [{\citenamefont {Vamivakas}\ \emph {et~al.}(2010)\citenamefont
  {Vamivakas}, \citenamefont {Lu}, \citenamefont {Matthiesen}, \citenamefont
  {Zhao}, \citenamefont {F{\"a}lt}, \citenamefont {Badolato},\ and\
  \citenamefont {Atat{\"u}re}}]{vamivakas_2010}%
  \BibitemOpen
  \bibfield  {author} {\bibinfo {author} {\bibfnamefont {A.~N.}\ \bibnamefont
  {Vamivakas}}, \bibinfo {author} {\bibfnamefont {C.~Y.}\ \bibnamefont {Lu}},
  \bibinfo {author} {\bibfnamefont {C.}~\bibnamefont {Matthiesen}}, \bibinfo
  {author} {\bibfnamefont {Y.}~\bibnamefont {Zhao}}, \bibinfo {author}
  {\bibfnamefont {S.}~\bibnamefont {F{\"a}lt}}, \bibinfo {author}
  {\bibfnamefont {A.}~\bibnamefont {Badolato}},\ and\ \bibinfo {author}
  {\bibfnamefont {M.}~\bibnamefont {Atat{\"u}re}},\ }\bibfield  {title}
  {\bibinfo {title} {Observation of spin-dependent quantum jumps via quantum
  dot resonance fluorescence},\ }\href {https://doi.org/10.1038/nature09359}
  {\bibfield  {journal} {\bibinfo  {journal} {Nature}\ }\textbf {\bibinfo
  {volume} {467}},\ \bibinfo {pages} {297} (\bibinfo {year}
  {2010})}\BibitemShut {NoStop}%
\bibitem [{\citenamefont {Matthiesen}\ \emph {et~al.}(2013)\citenamefont
  {Matthiesen}, \citenamefont {Geller}, \citenamefont {Schulte}, \citenamefont
  {Le~Gall}, \citenamefont {Hansom}, \citenamefont {Li}, \citenamefont
  {Hugues}, \citenamefont {Clarke},\ and\ \citenamefont
  {Atat{\"u}re}}]{matthiesen_2013}%
  \BibitemOpen
  \bibfield  {author} {\bibinfo {author} {\bibfnamefont {C.}~\bibnamefont
  {Matthiesen}}, \bibinfo {author} {\bibfnamefont {M.}~\bibnamefont {Geller}},
  \bibinfo {author} {\bibfnamefont {C.~H.~H.}\ \bibnamefont {Schulte}},
  \bibinfo {author} {\bibfnamefont {C.}~\bibnamefont {Le~Gall}}, \bibinfo
  {author} {\bibfnamefont {J.}~\bibnamefont {Hansom}}, \bibinfo {author}
  {\bibfnamefont {Z.}~\bibnamefont {Li}}, \bibinfo {author} {\bibfnamefont
  {M.}~\bibnamefont {Hugues}}, \bibinfo {author} {\bibfnamefont
  {E.}~\bibnamefont {Clarke}},\ and\ \bibinfo {author} {\bibfnamefont
  {M.}~\bibnamefont {Atat{\"u}re}},\ }\bibfield  {title} {\bibinfo {title}
  {Phase-locked indistinguishable photons with synthesized waveforms from a
  solid-state source},\ }\href {https://doi.org/10.1038/ncomms2601} {\bibfield
  {journal} {\bibinfo  {journal} {Nat. Commun.}\ }\textbf {\bibinfo {volume}
  {4}},\ \bibinfo {pages} {1600} (\bibinfo {year} {2013})}\BibitemShut
  {NoStop}%
\bibitem [{\citenamefont {Kurzmann}\ \emph
  {et~al.}(2016{\natexlab{b}})\citenamefont {Kurzmann}, \citenamefont {Merkel},
  \citenamefont {Labud}, \citenamefont {Ludwig}, \citenamefont {Wieck},
  \citenamefont {Lorke},\ and\ \citenamefont {Geller}}]{kurzmann_2016}%
  \BibitemOpen
  \bibfield  {author} {\bibinfo {author} {\bibfnamefont {A.}~\bibnamefont
  {Kurzmann}}, \bibinfo {author} {\bibfnamefont {B.}~\bibnamefont {Merkel}},
  \bibinfo {author} {\bibfnamefont {P.~A.}\ \bibnamefont {Labud}}, \bibinfo
  {author} {\bibfnamefont {A.}~\bibnamefont {Ludwig}}, \bibinfo {author}
  {\bibfnamefont {A.~D.}\ \bibnamefont {Wieck}}, \bibinfo {author}
  {\bibfnamefont {A.}~\bibnamefont {Lorke}},\ and\ \bibinfo {author}
  {\bibfnamefont {M.}~\bibnamefont {Geller}},\ }\bibfield  {title} {\bibinfo
  {title} {Optical {B}locking of {E}lectron {T}unneling into a {S}ingle
  {S}elf-{A}ssembled {Q}uantum {D}ot},\ }\href
  {https://doi.org/10.1103/PhysRevLett.117.017401} {\bibfield  {journal}
  {\bibinfo  {journal} {Phys. Rev. Lett.}\ }\textbf {\bibinfo {volume} {117}},\
  \bibinfo {pages} {017401} (\bibinfo {year} {2016}{\natexlab{b}})}\BibitemShut
  {NoStop}%
\bibitem [{\citenamefont {Lochner}\ \emph {et~al.}(2019)\citenamefont
  {Lochner}, \citenamefont {Kurzmann}, \citenamefont {Schott}, \citenamefont
  {Wieck}, \citenamefont {Ludwig}, \citenamefont {Lorke},\ and\ \citenamefont
  {Geller}}]{lochner_2019}%
  \BibitemOpen
  \bibfield  {author} {\bibinfo {author} {\bibfnamefont {P.}~\bibnamefont
  {Lochner}}, \bibinfo {author} {\bibfnamefont {A.}~\bibnamefont {Kurzmann}},
  \bibinfo {author} {\bibfnamefont {R.}~\bibnamefont {Schott}}, \bibinfo
  {author} {\bibfnamefont {A.~D.}\ \bibnamefont {Wieck}}, \bibinfo {author}
  {\bibfnamefont {A.}~\bibnamefont {Ludwig}}, \bibinfo {author} {\bibfnamefont
  {A.}~\bibnamefont {Lorke}},\ and\ \bibinfo {author} {\bibfnamefont
  {M.}~\bibnamefont {Geller}},\ }\bibfield  {title} {\bibinfo {title} {Contrast
  of 83{\%} in reflection measurements on a single quantum dot},\ }\href
  {https://doi.org/10.1038/s41598-019-45259-z} {\bibfield  {journal} {\bibinfo
  {journal} {Sci. Rep.}\ }\textbf {\bibinfo {volume} {9}},\ \bibinfo {pages}
  {8817} (\bibinfo {year} {2019})}\BibitemShut {NoStop}%
\bibitem [{\citenamefont {Bayer}\ \emph {et~al.}(2002)\citenamefont {Bayer},
  \citenamefont {Ortner}, \citenamefont {Stern}, \citenamefont {Kuther},
  \citenamefont {Gorbunov}, \citenamefont {Forchel}, \citenamefont {Hawrylak},
  \citenamefont {Fafard}, \citenamefont {Hinzer}, \citenamefont {Reinecke},
  \citenamefont {Walck}, \citenamefont {Reithmaier}, \citenamefont {Klopf},\
  and\ \citenamefont {Sch\"afer}}]{bayer_2002}%
  \BibitemOpen
  \bibfield  {author} {\bibinfo {author} {\bibfnamefont {M.}~\bibnamefont
  {Bayer}}, \bibinfo {author} {\bibfnamefont {G.}~\bibnamefont {Ortner}},
  \bibinfo {author} {\bibfnamefont {O.}~\bibnamefont {Stern}}, \bibinfo
  {author} {\bibfnamefont {A.}~\bibnamefont {Kuther}}, \bibinfo {author}
  {\bibfnamefont {A.~A.}\ \bibnamefont {Gorbunov}}, \bibinfo {author}
  {\bibfnamefont {A.}~\bibnamefont {Forchel}}, \bibinfo {author} {\bibfnamefont
  {P.}~\bibnamefont {Hawrylak}}, \bibinfo {author} {\bibfnamefont
  {S.}~\bibnamefont {Fafard}}, \bibinfo {author} {\bibfnamefont
  {K.}~\bibnamefont {Hinzer}}, \bibinfo {author} {\bibfnamefont {T.~L.}\
  \bibnamefont {Reinecke}}, \bibinfo {author} {\bibfnamefont {S.~N.}\
  \bibnamefont {Walck}}, \bibinfo {author} {\bibfnamefont {J.~P.}\ \bibnamefont
  {Reithmaier}}, \bibinfo {author} {\bibfnamefont {F.}~\bibnamefont {Klopf}},\
  and\ \bibinfo {author} {\bibfnamefont {F.}~\bibnamefont {Sch\"afer}},\
  }\bibfield  {title} {\bibinfo {title} {Fine structure of neutral and charged
  excitons in self-assembled
  {$\mathrm{In}(\mathrm{Ga})\mathrm{As}/(\mathrm{Al})\mathrm{GaAs}$} quantum
  dots},\ }\href {https://doi.org/10.1103/PhysRevB.65.195315} {\bibfield
  {journal} {\bibinfo  {journal} {Phys. Rev. B}\ }\textbf {\bibinfo {volume}
  {65}},\ \bibinfo {pages} {195315} (\bibinfo {year} {2002})}\BibitemShut
  {NoStop}%
\bibitem [{\citenamefont {Kroutvar}\ \emph {et~al.}(2004)\citenamefont
  {Kroutvar}, \citenamefont {Ducommun}, \citenamefont {Heiss}, \citenamefont
  {Bichler}, \citenamefont {Schuh}, \citenamefont {Abstreiter},\ and\
  \citenamefont {Finley}}]{kroutvar_2004}%
  \BibitemOpen
  \bibfield  {author} {\bibinfo {author} {\bibfnamefont {M.}~\bibnamefont
  {Kroutvar}}, \bibinfo {author} {\bibfnamefont {Y.}~\bibnamefont {Ducommun}},
  \bibinfo {author} {\bibfnamefont {D.}~\bibnamefont {Heiss}}, \bibinfo
  {author} {\bibfnamefont {M.}~\bibnamefont {Bichler}}, \bibinfo {author}
  {\bibfnamefont {D.}~\bibnamefont {Schuh}}, \bibinfo {author} {\bibfnamefont
  {G.}~\bibnamefont {Abstreiter}},\ and\ \bibinfo {author} {\bibfnamefont
  {J.~J.}\ \bibnamefont {Finley}},\ }\bibfield  {title} {\bibinfo {title}
  {Optically programmable electron spin memory using semiconductor quantum
  dots},\ }\href {https://doi.org/10.1038/nature03008} {\bibfield  {journal}
  {\bibinfo  {journal} {Nature}\ }\textbf {\bibinfo {volume} {432}},\ \bibinfo
  {pages} {81} (\bibinfo {year} {2004})}\BibitemShut {NoStop}%
\bibitem [{\citenamefont {Nesbet}(1971)}]{nesbet_1971}%
  \BibitemOpen
  \bibfield  {author} {\bibinfo {author} {\bibfnamefont {R.~K.}\ \bibnamefont
  {Nesbet}},\ }\bibfield  {title} {\bibinfo {title} {Where {S}emiclassical
  {R}adiation {T}heory {F}ails},\ }\href
  {https://doi.org/10.1103/PhysRevLett.27.553} {\bibfield  {journal} {\bibinfo
  {journal} {Phys. Rev. Lett.}\ }\textbf {\bibinfo {volume} {27}},\ \bibinfo
  {pages} {553} (\bibinfo {year} {1971})}\BibitemShut {NoStop}%
\bibitem [{\citenamefont {Debus}\ \emph
  {et~al.}(2014{\natexlab{b}})\citenamefont {Debus}, \citenamefont
  {Shamirzaev}, \citenamefont {Dunker}, \citenamefont {Sapega}, \citenamefont
  {Ivchenko}, \citenamefont {Yakovlev}, \citenamefont {Toropov},\ and\
  \citenamefont {Bayer}}]{debus_2014}%
  \BibitemOpen
  \bibfield  {author} {\bibinfo {author} {\bibfnamefont {J.}~\bibnamefont
  {Debus}}, \bibinfo {author} {\bibfnamefont {T.~S.}\ \bibnamefont
  {Shamirzaev}}, \bibinfo {author} {\bibfnamefont {D.}~\bibnamefont {Dunker}},
  \bibinfo {author} {\bibfnamefont {V.~F.}\ \bibnamefont {Sapega}}, \bibinfo
  {author} {\bibfnamefont {E.~L.}\ \bibnamefont {Ivchenko}}, \bibinfo {author}
  {\bibfnamefont {D.~R.}\ \bibnamefont {Yakovlev}}, \bibinfo {author}
  {\bibfnamefont {A.~I.}\ \bibnamefont {Toropov}},\ and\ \bibinfo {author}
  {\bibfnamefont {M.}~\bibnamefont {Bayer}},\ }\bibfield  {title} {\bibinfo
  {title} {{Spin-flip Raman scattering of the
  $\ensuremath{\Gamma}\text{\ensuremath{-}}X$ mixed exciton in indirect band
  gap (In,Al)As/AlAs quantum dots}},\ }\href
  {https://doi.org/10.1103/PhysRevB.90.125431} {\bibfield  {journal} {\bibinfo
  {journal} {Phys. Rev. B}\ }\textbf {\bibinfo {volume} {90}},\ \bibinfo
  {pages} {125431} (\bibinfo {year} {2014}{\natexlab{b}})}\BibitemShut
  {NoStop}%
\bibitem [{\citenamefont {Shabaev}\ \emph {et~al.}(2003)\citenamefont
  {Shabaev}, \citenamefont {Efros}, \citenamefont {Gammon},\ and\ \citenamefont
  {Merkulov}}]{shabaev_2003}%
  \BibitemOpen
  \bibfield  {author} {\bibinfo {author} {\bibfnamefont {A.}~\bibnamefont
  {Shabaev}}, \bibinfo {author} {\bibfnamefont {A.~L.}\ \bibnamefont {Efros}},
  \bibinfo {author} {\bibfnamefont {D.}~\bibnamefont {Gammon}},\ and\ \bibinfo
  {author} {\bibfnamefont {I.~A.}\ \bibnamefont {Merkulov}},\ }\bibfield
  {title} {\bibinfo {title} {Optical readout and initialization of an electron
  spin in a single quantum dot},\ }\href
  {https://doi.org/10.1103/PhysRevB.68.201305} {\bibfield  {journal} {\bibinfo
  {journal} {Phys. Rev. B}\ }\textbf {\bibinfo {volume} {68}},\ \bibinfo
  {pages} {201305(R)} (\bibinfo {year} {2003})}\BibitemShut {NoStop}%
\bibitem [{\citenamefont {Calarco}\ \emph {et~al.}(2003)\citenamefont
  {Calarco}, \citenamefont {Datta}, \citenamefont {Fedichev}, \citenamefont
  {Pazy},\ and\ \citenamefont {Zoller}}]{calarco_2003}%
  \BibitemOpen
  \bibfield  {author} {\bibinfo {author} {\bibfnamefont {T.}~\bibnamefont
  {Calarco}}, \bibinfo {author} {\bibfnamefont {A.}~\bibnamefont {Datta}},
  \bibinfo {author} {\bibfnamefont {P.}~\bibnamefont {Fedichev}}, \bibinfo
  {author} {\bibfnamefont {E.}~\bibnamefont {Pazy}},\ and\ \bibinfo {author}
  {\bibfnamefont {P.}~\bibnamefont {Zoller}},\ }\bibfield  {title} {\bibinfo
  {title} {Spin-based all-optical quantum computation with quantum dots:
  {U}nderstanding and suppressing decoherence},\ }\href
  {https://doi.org/10.1103/PhysRevA.68.012310} {\bibfield  {journal} {\bibinfo
  {journal} {Phys. Rev. A}\ }\textbf {\bibinfo {volume} {68}},\ \bibinfo
  {pages} {012310} (\bibinfo {year} {2003})}\BibitemShut {NoStop}%
\bibitem [{\citenamefont {Lu}\ \emph {et~al.}(2010)\citenamefont {Lu},
  \citenamefont {Zhao}, \citenamefont {Vamivakas}, \citenamefont {Matthiesen},
  \citenamefont {F\"alt}, \citenamefont {Badolato},\ and\ \citenamefont
  {Atat\"ure}}]{lu_2010}%
  \BibitemOpen
  \bibfield  {author} {\bibinfo {author} {\bibfnamefont {C.-Y.}\ \bibnamefont
  {Lu}}, \bibinfo {author} {\bibfnamefont {Y.}~\bibnamefont {Zhao}}, \bibinfo
  {author} {\bibfnamefont {A.~N.}\ \bibnamefont {Vamivakas}}, \bibinfo {author}
  {\bibfnamefont {C.}~\bibnamefont {Matthiesen}}, \bibinfo {author}
  {\bibfnamefont {S.}~\bibnamefont {F\"alt}}, \bibinfo {author} {\bibfnamefont
  {A.}~\bibnamefont {Badolato}},\ and\ \bibinfo {author} {\bibfnamefont
  {M.}~\bibnamefont {Atat\"ure}},\ }\bibfield  {title} {\bibinfo {title}
  {Direct measurement of spin dynamics in {$\mathrm{InAs}/\mathrm{GaAs}$}
  quantum dots using time-resolved resonance fluorescence},\ }\href
  {https://doi.org/10.1103/PhysRevB.81.035332} {\bibfield  {journal} {\bibinfo
  {journal} {Phys. Rev. B}\ }\textbf {\bibinfo {volume} {81}},\ \bibinfo
  {pages} {035332} (\bibinfo {year} {2010})}\BibitemShut {NoStop}%
\bibitem [{\citenamefont {Hanson}\ \emph {et~al.}(2007)\citenamefont {Hanson},
  \citenamefont {Kouwenhoven}, \citenamefont {Petta}, \citenamefont {Tarucha},\
  and\ \citenamefont {Vandersypen}}]{hanson_2007}%
  \BibitemOpen
  \bibfield  {author} {\bibinfo {author} {\bibfnamefont {R.}~\bibnamefont
  {Hanson}}, \bibinfo {author} {\bibfnamefont {L.~P.}\ \bibnamefont
  {Kouwenhoven}}, \bibinfo {author} {\bibfnamefont {J.~R.}\ \bibnamefont
  {Petta}}, \bibinfo {author} {\bibfnamefont {S.}~\bibnamefont {Tarucha}},\
  and\ \bibinfo {author} {\bibfnamefont {L.~M.~K.}\ \bibnamefont
  {Vandersypen}},\ }\bibfield  {title} {\bibinfo {title} {Spins in few-electron
  quantum dots},\ }\href {https://doi.org/10.1103/RevModPhys.79.1217}
  {\bibfield  {journal} {\bibinfo  {journal} {Rev. Mod. Phys.}\ }\textbf
  {\bibinfo {volume} {79}},\ \bibinfo {pages} {1217} (\bibinfo {year}
  {2007})}\BibitemShut {NoStop}%
\bibitem [{\citenamefont {Klimov}\ \emph {et~al.}(2000)\citenamefont {Klimov},
  \citenamefont {Mikhailovsky}, \citenamefont {Xu}, \citenamefont {Malko},
  \citenamefont {Hollingsworth}, \citenamefont {Leatherdale}, \citenamefont
  {Eisler},\ and\ \citenamefont {Bawendi}}]{klimov_2000}%
  \BibitemOpen
  \bibfield  {author} {\bibinfo {author} {\bibfnamefont {V.~I.}\ \bibnamefont
  {Klimov}}, \bibinfo {author} {\bibfnamefont {A.~A.}\ \bibnamefont
  {Mikhailovsky}}, \bibinfo {author} {\bibfnamefont {S.}~\bibnamefont {Xu}},
  \bibinfo {author} {\bibfnamefont {A.}~\bibnamefont {Malko}}, \bibinfo
  {author} {\bibfnamefont {J.~A.}\ \bibnamefont {Hollingsworth}}, \bibinfo
  {author} {\bibfnamefont {C.~A.}\ \bibnamefont {Leatherdale}}, \bibinfo
  {author} {\bibfnamefont {H.-J.}\ \bibnamefont {Eisler}},\ and\ \bibinfo
  {author} {\bibfnamefont {M.~G.}\ \bibnamefont {Bawendi}},\ }\bibfield
  {title} {\bibinfo {title} {Optical {G}ain and {S}timulated {E}mission in
  {N}anocrystal {Q}uantum {D}ots},\ }\href
  {https://doi.org/10.1126/science.290.5490.314} {\bibfield  {journal}
  {\bibinfo  {journal} {Science}\ }\textbf {\bibinfo {volume} {290}},\ \bibinfo
  {pages} {314} (\bibinfo {year} {2000})}\BibitemShut {NoStop}%
\bibitem [{\citenamefont {Pietryga}\ \emph {et~al.}(2008)\citenamefont
  {Pietryga}, \citenamefont {Zhuravlev}, \citenamefont {Whitehead},
  \citenamefont {Klimov},\ and\ \citenamefont {Schaller}}]{pietryga_2008}%
  \BibitemOpen
  \bibfield  {author} {\bibinfo {author} {\bibfnamefont {J.~M.}\ \bibnamefont
  {Pietryga}}, \bibinfo {author} {\bibfnamefont {K.~K.}\ \bibnamefont
  {Zhuravlev}}, \bibinfo {author} {\bibfnamefont {M.}~\bibnamefont
  {Whitehead}}, \bibinfo {author} {\bibfnamefont {V.~I.}\ \bibnamefont
  {Klimov}},\ and\ \bibinfo {author} {\bibfnamefont {R.~D.}\ \bibnamefont
  {Schaller}},\ }\bibfield  {title} {\bibinfo {title} {Evidence for
  {B}arrierless {A}uger {R}ecombination in {$\mathrm{PbSe}$} {N}anocrystals:
  {A} {P}ressure-{D}ependent {S}tudy of {T}ransient {O}ptical {A}bsorption},\
  }\href {https://doi.org/10.1103/PhysRevLett.101.217401} {\bibfield  {journal}
  {\bibinfo  {journal} {Phys. Rev. Lett.}\ }\textbf {\bibinfo {volume} {101}},\
  \bibinfo {pages} {217401} (\bibinfo {year} {2008})}\BibitemShut {NoStop}%
\bibitem [{\citenamefont {Kerski}\ \emph {et~al.}(2023)\citenamefont {Kerski},
  \citenamefont {Mannel}, \citenamefont {Lochner}, \citenamefont
  {Kleinherbers}, \citenamefont {Kurzmann}, \citenamefont {Ludwig},
  \citenamefont {Wieck}, \citenamefont {K{\"o}nig}, \citenamefont {Lorke},\
  and\ \citenamefont {Geller}}]{kerski_2023}%
  \BibitemOpen
  \bibfield  {author} {\bibinfo {author} {\bibfnamefont {J.}~\bibnamefont
  {Kerski}}, \bibinfo {author} {\bibfnamefont {H.}~\bibnamefont {Mannel}},
  \bibinfo {author} {\bibfnamefont {P.}~\bibnamefont {Lochner}}, \bibinfo
  {author} {\bibfnamefont {E.}~\bibnamefont {Kleinherbers}}, \bibinfo {author}
  {\bibfnamefont {A.}~\bibnamefont {Kurzmann}}, \bibinfo {author}
  {\bibfnamefont {A.}~\bibnamefont {Ludwig}}, \bibinfo {author} {\bibfnamefont
  {A.~D.}\ \bibnamefont {Wieck}}, \bibinfo {author} {\bibfnamefont
  {J.}~\bibnamefont {K{\"o}nig}}, \bibinfo {author} {\bibfnamefont
  {A.}~\bibnamefont {Lorke}},\ and\ \bibinfo {author} {\bibfnamefont
  {M.}~\bibnamefont {Geller}},\ }\bibfield  {title} {\bibinfo {title}
  {Post-processing of real-time quantum event measurements for an optimal
  bandwidth},\ }\href {https://doi.org/10.1038/s41598-023-28273-0} {\bibfield
  {journal} {\bibinfo  {journal} {Sci. Rep.}\ }\textbf {\bibinfo {volume}
  {13}},\ \bibinfo {pages} {1105} (\bibinfo {year} {2023})}\BibitemShut
  {NoStop}%
\bibitem [{\citenamefont {Landauer}(1998)}]{landauer_1998}%
  \BibitemOpen
  \bibfield  {author} {\bibinfo {author} {\bibfnamefont {R.}~\bibnamefont
  {Landauer}},\ }\bibfield  {title} {\bibinfo {title} {The noise is the
  signal},\ }\href {https://doi.org/10.1038/33551} {\bibfield  {journal}
  {\bibinfo  {journal} {Nature}\ }\textbf {\bibinfo {volume} {392}},\ \bibinfo
  {pages} {658} (\bibinfo {year} {1998})}\BibitemShut {NoStop}%
\bibitem [{\citenamefont {Smith}(1995)}]{smith_1995}%
  \BibitemOpen
  \bibfield  {author} {\bibinfo {author} {\bibfnamefont {P.~J.}\ \bibnamefont
  {Smith}},\ }\bibfield  {title} {\bibinfo {title} {A {R}ecursive {F}ormulation
  of the {O}ld {P}roblem of {O}btaining {M}oments from {C}umulants and {V}ice
  {V}ersa},\ }\href {https://doi.org/10.1080/00031305.1995.10476146} {\bibfield
   {journal} {\bibinfo  {journal} {Am. Stat.}\ }\textbf {\bibinfo {volume}
  {49}},\ \bibinfo {pages} {217} (\bibinfo {year} {1995})}\BibitemShut
  {NoStop}%
\bibitem [{\citenamefont {Johnson}\ \emph {et~al.}(2005)\citenamefont
  {Johnson}, \citenamefont {Kemp},\ and\ \citenamefont
  {Kotz}}]{johnson_univariate_2005}%
  \BibitemOpen
  \bibfield  {author} {\bibinfo {author} {\bibfnamefont {N.~L.}\ \bibnamefont
  {Johnson}}, \bibinfo {author} {\bibfnamefont {A.~W.}\ \bibnamefont {Kemp}},\
  and\ \bibinfo {author} {\bibfnamefont {S.}~\bibnamefont {Kotz}},\ }\href@noop
  {} {\emph {\bibinfo {title} {Univariate {D}iscrete {D}istributions}}}\
  (\bibinfo  {publisher} {Wiley},\ \bibinfo {address} {Hoboken},\ \bibinfo
  {year} {2005})\BibitemShut {NoStop}%
\end{thebibliography}%

\end{document}